\documentclass[aps,prd,preprint,superscriptaddress,showpacs]{revtex4}

\usepackage{graphicx}
\usepackage{dcolumn}
\usepackage{bm}
\usepackage{epsfig}
\begin{document}

\preprint{\vbox{\hbox{\hfil CLNS 04/1860}
                        \hbox{\hfil CLEO 04-02} }} 
\title{Measurement of the $B$-Meson Inclusive Semileptonic Branching
  Fraction and Electron-Energy Moments} 

\author{A.~H.~Mahmood}
\affiliation{University of Texas - Pan American, Edinburg, Texas 78539}
\author{S.~E.~Csorna}
\affiliation{Vanderbilt University, Nashville, Tennessee 37235}
\author{G.~Bonvicini}
\author{D.~Cinabro}
\author{M.~Dubrovin}
\affiliation{Wayne State University, Detroit, Michigan 48202}
\author{A.~Bornheim}
\author{E.~Lipeles}
\author{S.~P.~Pappas}
\author{A.~Shapiro}
\author{A.~J.~Weinstein}
\affiliation{California Institute of Technology, Pasadena, California 91125}
% ...
\collaboration{CLEO Collaboration} 
\noaffiliation

\date{\today}

\begin{abstract}

We report a new measurement of the $B$-meson semileptonic decay momentum
spectrum that has been made with a sample of 9.4 fb$^{-1}$ of $e^+e^-$
data collected with the CLEO~II detector at the $\Upsilon(4S)$
resonance.  Electrons from primary
semileptonic decays and secondary charm decays were separated by using
charge and angular correlations in $\Upsilon(4S)$ events with a
high-momentum lepton and an additional electron.  We determined the
semileptonic branching fraction to be ${\mathcal B}(B \rightarrow X e^+
\nu_e) = (10.91 \pm 0.09 \pm 0.24)\%$ from the normalization of the
electron-energy spectrum.  We also measured the moments of the electron
energy spectrum with minimum energies from 0.6 GeV to 1.5 GeV.
\end{abstract}



\maketitle


\section{Introduction}
\label{sec:Intro}
Semileptonic decays of $B$ mesons have been the principal tool for
determining the Cabibbo-Kobayashi-Maskawa (CKM) matrix elements
$V_{cb}$ and $V_{ub}$ that govern the weak-current couplings of $b$
quarks through external $W^{\pm}$ emission. 
This reliance results from the inherent simplicity of semileptonic
decays, which render more direct access to the underlying quark
couplings than do hadronic decays.  Nonperturbative hadronic effects
play a significant role in the details of semileptonic $B$ decays,
however, and pose considerable challenges to the interpretation of
precision inclusive and exclusive measurements.  This has been
demonstrated by puzzles such as a measured $B$ semileptonic branching
fraction that has been persistently smaller than theoretical
expectations~\cite{Bigi:1994fm, Albrecht:1993pu, Barish:1995cx, Aubert:2002uf,
Abe:2002du}. 

In recent years, Heavy Quark Effective Theory (HQET) has emerged as a
powerful tool in the interpretation of the properties of mesons
containing a heavy quark.  Rooted in
QCD and implemented through the Operator Product Expansion (OPE), HQET
provides a rigorous procedure for expressing the observables of
semileptonic and rare $B$ decays as expansions in perturbative and
non-perturbative parameters~\cite{Shifman:1985wx, Chay:1990da,
Bigi:1992su, Bigi:1993fe, Adel:1994ah}.  If the validity of this
formulation of QCD can be demonstrated by detailed comparison with data,
then HQET/OPE can be used to extract the CKM parameter $|V_{cb}|$ from
the $B$ semileptonic branching fraction and lifetime with uncertainties
that are significantly reduced.  

Voloshin first suggested that the moments of the lepton-energy
spectrum in inclusively measured semileptonic $B$ decays could provide precise
information about the quark-mass difference $m_b -
m_c$~\cite{Voloshin:1995cy}.  A succession of authors have expanded on
this proposal to include moments of other observables of semileptonic
decays and the electromagnetic penguin decay $B \rightarrow X_s
\gamma$~\cite{Gremm:1996yn, Neubert:1994um}.  Measurements have been
presented by the CLEO~\cite{Chen:2001fj, Cronin-Hennessy:2001fk} and
DELPHI~\cite{Calvi:2002wc} collaborations.  Recently, there have been
efforts to provide a consistent framework for the interpretation of
these measurements.  Battaglia {\it et al.}~\cite{Battaglia:2002tm} have
performed fits to order $1/m_b^3$ of the preliminary moment measurements
of the DELPHI collaboration.  Bauer, Ligeti, Luke, and Manohar have
presented expressions for various moments of inclusive $B$ decay to
order $\alpha_s^2 \beta_0$ and $\Lambda^3_{QCD}$ for several mass
schemes~\cite{Bauer:2002sh}.  Fits to the moments of different
distributions and to measurements that sample different regions of phase
space serve as checks of the overall validity of the HQET/OPE approach.
In particular, such tests probe for potential violations of the
underlying assumption of quark-hadron duality.  

In this paper we present a new measurement of inclusive semileptonic $B$
decays that has been made with the complete data sample obtained with
the CLEO~II detector at the Cornell Electron Storage Ring (CESR).  The
momentum spectrum for primary semileptonic decays  $B \rightarrow X e
\nu$ was isolated through the use of charge and angular correlations in
$\Upsilon(4S) \rightarrow B {\bar B}$ dilepton events.  The technique of
using angular correlations in events with a high-momentum lepton was
first used by CLEO for measurements of $B$ decays to
kaons~\cite{tipton_thesis}.  It was subsequently applied to measurements
of semileptonic $B$ decays by ARGUS~\cite{Albrecht:1993pu} and
CLEO~\cite{Barish:1995cx}.  In this paper we use the normalization of
the measured electron-momentum spectrum to obtain the $B$ semileptonic
branching fraction and the detailed shape of the spectrum to measure the
electron-energy moments with various minimum-energy cuts.  The results presented
here supersede the previous CLEO~II measurement of the semileptonic
branching fraction~\cite{Barish:1995cx}, which was based on the first
fifth of the CLEO~II data sample.  This paper presents an initial
interpretation of the electron-energy moments in the context of HQET.  A
forthcoming publication~\cite{mother_of_all_moments} will provide a
comprehensive interpretation of 
these measurements and other moments of inclusive $B$ decays that have
previously been reported by CLEO~\cite{Lipeles_moments, Chen:2001fj}.   

\section{CLEO~II Detector and Event Sample}
\label{sec:detector}
The CLEO~II detector, which has since been replaced by the CLEO~III
detector, was a general purpose magnetic spectrometer with a 
1.5-T superconducting solenoidal magnet and excellent charged-particle
tracking and electromagnetic calorimetry.  Detailed descriptions of the
detector and its performance have been presented
previously~\cite{Kubota:1992ww, Hill:1998ea}.  Two configurations of the
detector were used to collect the data sample of this paper.  The first
third of the data was obtained with a tracking system that consisted of
three concentric cylindrical drift chambers surrounding the beam line.
The remaining two thirds were collected after an upgrade that included
the replacement of the innermost straw-tube drift chamber with a three-layer
silicon vertex detector and a change of the gas mixture from
argon-ethane to helium-propane in the main drift chamber.  The
tracking system provided solid-angle coverage of 95\% of $4\pi$ in both
configurations, and the momentum resolution at 2~GeV/$c$ was
0.6\%.  The tracking devices also provided specific-ionization
measurements for hadron identification, with additional $\pi/K/p$
discrimination provided by a time-of-flight scintillator system located
just beyond the tracking.  The final detector system inside the
solenoidal magnet was a 7800-crystal CsI (Tl) electromagnetic
calorimeter with solid-angle coverage of 98\% of $4 \pi$.  The
calorimeter was crucial for electron identification and 
provided excellent efficiency and energy resolution for photons,
yielding a typical mass resolution for $\pi^0$ reconstruction of 6~MeV (FWHM). 
The outermost detector component was the muon identification system,
which consisted of layers of proportional-tube chambers embedded at
three depths in the iron flux return surrounding the magnet.  

The $B$-meson sample for this analysis was obtained by selecting
multihadronic events from 9.4~fb$^{-1}$ of CESR $e^+e^-$ annihilation
data at 10.58~GeV, the peak of the $\Upsilon(4S)$ resonance.  A
requirement of at least five well-reconstructed charged tracks was
imposed to suppress low-multiplicity background processes: $\tau$-pair,
radiative Bhabha, radiative $\mu$-pair, and two-photon events.  
Contributions from continuum events $e^+e^- \rightarrow q {\bar q}$ 
($q$=$d$, $u$, $s$, or $c$) were determined with 4.5~fb$^{-1}$ of data 
collected at a center-of-mass energy approximately 60~MeV below the 
$\Upsilon(4S)$, where there is no production of $B {\bar B}$.  Before 
subtraction, below-resonance distributions were scaled to account for 
the difference in the integrated luminosities of the two samples and for 
the 1/$s$ dependence of the $e^+e^- \rightarrow q {\bar q}$ cross section.  
The scale factor was computed with measured integrated luminosities and CESR
beam energies, and confirmed by direct determination of the
on-resonance/below-resonance ratio of charged-track yields above the
kinematic limit for the momenta of $B$-decay daughters at the
$\Upsilon(4S)$.  These independent 
determinations agreed within approximately 0.5\%, and a 1\% systematic
uncertainty in the correction was assumed.  
The $\Upsilon(4S)$ sample was determined to include 9.7 million $B {\bar
  B}$ events.   

\section{Selection of Dilepton Events}
\label{sec:event_selection}
For the measurement of the inclusive electron spectrum in semileptonic
$B$ decay, we selected events with a high-momentum (tag) lepton. 
The tag lepton could be either an electron or 
a muon, and was required to have a minimum momentum of  1.4~GeV/$c$ and
a maximum momentum of 2.6~GeV/$c$.  Such leptons are predominantly
produced  in the semileptonic decay of one of the two $B$ mesons in an
$\Upsilon(4S)$ decay.  In events with tags, we searched for an
accompanying (signal) electron, with minimum momentum 0.6~GeV/$c$.  These
electrons were primarily from the semileptonic decay of the other $B$
meson or from semileptonic decay of a charmed daughter of either the
same or the other $B$ meson.  The procedure for disentangling these
components is described in Section~\ref{sec:spectrum}.  

All identified leptons were required to project into the central part of
the detector ($|\cos \theta|<0.71$, where $\theta$ is the angle between
the lepton direction and the beam axis).  This fiducial requirement
ensured the most reliable and best-understood track reconstruction and
lepton identification.  Requirements on tracking residuals, impact
parameters, and the fraction of tracking layers traversed that had high-quality
hits provided additional assurance of reliably determined momenta.  

Muons were identified by their ability to penetrate detector material
and register hits in the muon chambers.  Accepted muon tags were
required to reach a depth of at least five nuclear interaction lengths
and to have the expected corroborating hits at smaller depths.  The
efficiency for detecting muons was greater than 90\%, and the
probability for a hadron track to be misidentified as a muon was less
than 1\%.  Because muons were used only as tags in this analysis, the
results are quite insensitive to the details of muon identification.  

Electrons were selected with criteria that relied mostly on the ratio of
the energy deposited in the electromagnetic calorimeter to the measured
momentum ($E/p$) and on the specific ionization ($dE/dx$) measured in
the tracking chambers.  The measurement of the $B \rightarrow X e \nu$
signal spectrum is very sensitive to the details of electron
identification; this was the dominant systematic uncertainty in our
previous measurement of the $B \rightarrow X e \nu$
spectrum~\cite{Albrecht:1993pu}.  For this reason, we developed a
customized electron-identification procedure for this analysis and have
made extensive studies of efficiencies and misidentification rates.  

The standard CLEO II electron-identification procedure was a likelihood-based
selection that combined measurements of $dE/dx$, time-of-flight, and
calorimeter information including $E/p$ and transverse shower shape.
The selection was trained and its efficiency and misidentification
probability were determined using data.  Electrons from radiative Bhabha
events, embedded in hadronic events, were used for the efficiency
measurement, and samples of tagged hadron tracks (pions from $K^0_S$
decays, kaons from $D^* \rightarrow D^0 \rightarrow K^- \pi^+$, and
$p/{\bar p}$ from $\Lambda/{\bar \Lambda}$ decays) were used to measure
misidentification rates.  This procedure provided highly optimized
electron identification, with efficiency ranging from 88\% at
0.6~GeV/$c$ to 93\% at 2.2~GeV/$c$, as well as hadron-misidentification
probabilities that were less than 0.1\% over nearly all of the momentum
range used for our spectrum measurement.

Detailed studies of the efficiency determination for this standard
electron identification revealed a bias in measurements made
with embedded radiative Bhabha events that could be significant for
precision measurements.  This appeared as a dip in the
efficiency beginning at $\sim 1.8$~GeV/$c$, which was traced to the
inclusion of shower-shape variables in the likelihood.  Some electrons
from radiative Bhabha events were lost because of distortion of the
electron shower due to overlap of the electron and the radiated photon.
While radiative Bhabha event-selection cuts were developed to mitigate
this effect, it was felt that the associated uncertainty in the momentum
dependence of the electron-identification procedure would be a
significant systematic limitation on our spectrum measurement.  Since
the background due to misidentified hadrons was judged to be negligible
at higher momenta, we developed an alternative procedure that sacrificed
some background rejection in favor of a more reliably determined
efficiency.  The new procedure used the full likelihood analysis below
1~GeV/$c$ and 
simple cuts on the key variables above 1~GeV/$c$: $E/p$ between 0.85 and
1.1 and measured $dE/dx$ no more than $2\sigma$ below the
expected value for an electron.  A time-of-flight requirement provided
additional hadron (primarily kaon) rejection between 1.0 and
1.6~GeV/$c$.  There was no requirement on shower shape above 1~GeV/$c$,
and the previously mentioned momentum-dependent bias was eliminated.  

We used several ``veto'' cuts to minimize backgrounds from sources other
than semileptonic decays.  We eliminated any tag or signal electron that
could be paired with another lepton of the same type and opposite charge
if the pair mass was within $3\sigma$ of the $J/\psi$
mass.  Monte Carlo simulations showed this veto to be approximately 58\%
efficient in rejecting electrons from $J/\psi$, while introducing an
inefficiency of 0.5\% into the selection of electrons from semileptonic
$B$ decays.  Electrons from $\pi^0$ Dalitz decays were rejected when the
three-body invariant mass of a combination of the candidate electron,
any oppositely charged track of momentum greater than 0.5~GeV/$c$ and a
photon was within $3\sigma$ of the $\pi^0$ mass.  In
this case, the efficiency for rejection was 29\% and the
inefficiency for semileptonic-decay electrons was less than 0.5\%.  Photon
conversions were rejected based on track-quality variables (e.g. the
distance of closest approach to the event vertex) and on the properties
and locations of vertices formed by pairing electron candidates with
oppositely charged tracks.  These criteria were found to be 56\% 
efficient in rejecting electrons from photon conversions and to contribute
an inefficiency for detecting electrons from $B \rightarrow X e \nu$ of
2\%.  For each of these vetoed processes, Monte Carlo simulations were used 
to estimate the background that ``leaked'' into our final sample, as is 
discussed in Sec.~\ref{sec:spectrum_systematics}.  

\section{Measurement of the Electron Momentum Spectra in Lepton-Tagged
  Events}
\label{sec:spectrum}

\subsection{Method}
\label{sec:spectrum_method}
The determination of the $B$-meson semileptonic branching fraction and
electron-energy moments demands a background-free sample of $B \rightarrow
X \ell \nu$ decays that covers as much of the available phase space as
possible.  The requirement of a lepton tag of minimum momentum
1.4~GeV/$c$ in $\Upsilon(4S) \rightarrow B {\bar B}$ events selects a
sample of semileptonic $B$ decays that is more than 97\% pure.  This
allows study of ``signal'' electron production from the other $B$ in the event with
small backgrounds and components that can be readily disentangled by
using charge and kinematic correlations.  In our analysis we searched
for signal electrons with momenta of at least 0.6~GeV/$c$.  This
minimum-momentum requirement was a compromise, allowing measurement of
approximately 94\% of the full $B$ semileptonic decay spectrum, while
excluding low-momentum electrons for which the systematic uncertainties
in efficiency determinations and hadronic backgrounds were significant.  

There are three main sources of signal electrons in lepton-tagged events,
summarized in Table~{\ref{tb:charge_correlations}.
\begin{table}
\begin{center}
\begin{tabular}{|l|c|c|}
\hline
& {\bf Unmixed Events} & {\bf Mixed Events} \\
\hline
Primary Events & $ \ell^{+} \leftarrow
\bar{b} \hspace{0.35 in} b \hspace{0.1 in} \longrightarrow  \hspace{0.1
in} e^{-} $ & $ \ell^{+} \leftarrow \bar{b}
\hspace{0.35 in} \bar{b} \hspace{0.1 in} \longrightarrow  \hspace{0.1 in} e^{+} $ \\
\hline
Opposite $B$ Secondary Events & $
\ell^{+} \leftarrow \bar{b} \hspace{0.35 in} b \to c \to
e^{+} $ & $\ell^{+} \leftarrow \bar{b} \hspace{0.35
in} \bar{b} \to \bar{c} \to e^{-} $ \\
\hline
Same $B$ Secondary Events & \multicolumn{2}{c|}{$\ell^{+} \leftarrow \bar{b}
  \longrightarrow \bar{c} \to e^{-} $}\\ 
\hline
\end{tabular}
\end{center}
\caption{\label{tb:charge_correlations}
  {Charge correlations for dilepton $B\bar{B}$ events.  The $\ell^+$
  denotes the tag lepton.}}
\end{table}
The key to discriminating among these sources is to measure the
spectra of signal electrons separately for events with a tag of the same charge and for
those with a tag of the opposite charge.  Semileptonic decay of the
other $B$ meson gives a signal electron with charge opposite to that of the
tag (if $B^0 {\bar B^0}$ mixing is ignored).  Semileptonic decay of a charm
meson that is a daughter of the other 
$B$ gives a signal electron of the same charge as the tag (again
ignoring $B^0 {\bar B^0}$ mixing).  Semileptonic decay
of a charm meson from the same $B$ gives a signal electron with the opposite charge
from the tag, but with a kinematic signature that makes its contribution
easy to isolate.  The effect of $B^0 {\bar B^0}$ mixing is to reverse
the charge correlations in a known proportion of events.  We use these
charge correlations to extract statistically the primary and secondary
spectra from the unlike-sign and like-sign spectra.  We assume that
charged and neutral $B$ mesons have the same decay rates and
lepton-energy spectra for primary semileptonic decays. 

Discrimination of same-$B$ signal electrons from opposite-$B$ signal
electrons in the unlike-sign sample relies on the kinematics of
production just above $B{\bar B}$ threshold.  At the $\Upsilon(4S)$, the
$B$ and the ${\bar B}$ are produced nearly at 
rest.  There is little correlation between the directions of a tag
lepton and of an accompanying electron if they are the daughters of
different $B$ mesons.  If they originate from the same $B$, however,
there is a strong tendency for the tag and the electron to be
back-to-back.  The correlation between the opening angle 
$\theta_{\ell e}$ of the tag lepton and the signal electron and the
signal electron momentum $p_e$ has been studied 
with Monte Carlo simulations of $B {\bar B}$ events and is illustrated 
in Fig.~\ref{fg:diag}.  
\begin{figure}[hbtp]
\centerline{
\begin{minipage}{8 cm}
\epsfxsize=8 cm
\epsfysize=8 cm
\epsffile{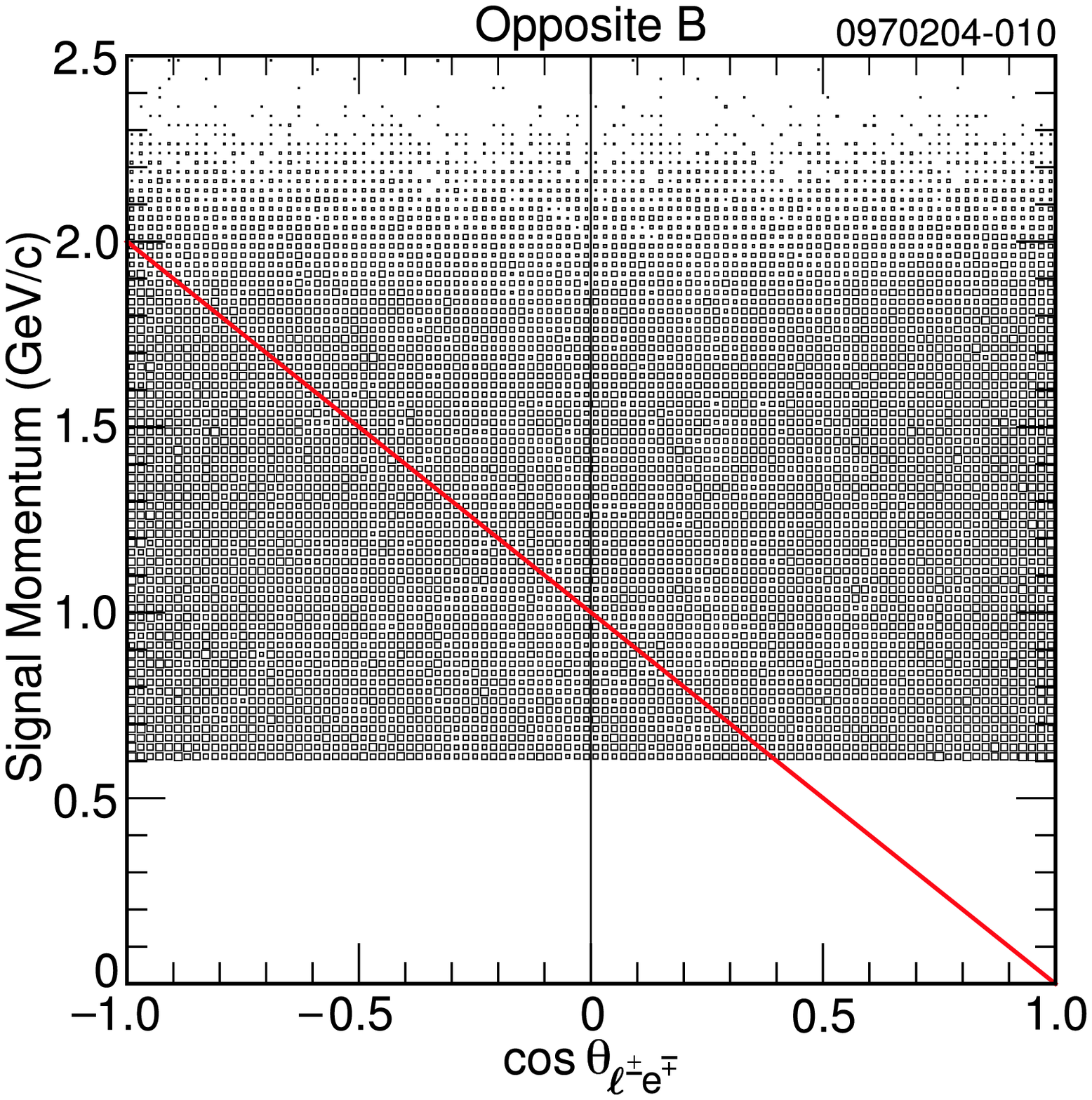}
\end{minipage}
\begin{minipage}{8 cm}
\epsfxsize=8 cm
\epsfysize=8 cm
\epsffile{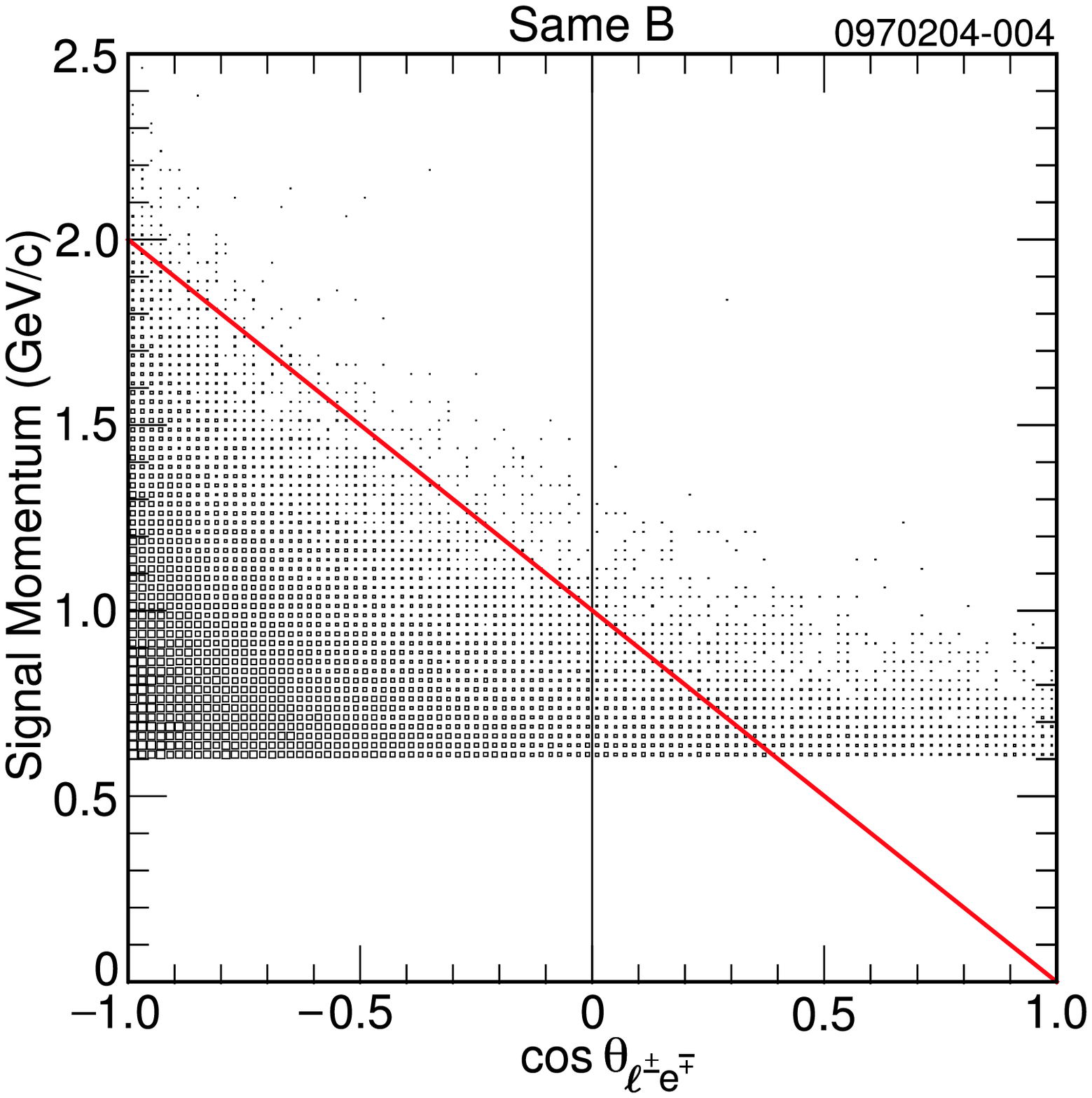}
\end{minipage}
} 
\caption[fg:diag]
{\label{fg:diag} Monte Carlo simulation of electron momentum versus the
cosine of the opening angle between the tag lepton and the signal
electron ($\cos\ \theta_{le}$) for unlike-sign dilepton pairs from opposite $B$'s
(top) and from the same $B$ (right).  The line indicates $p_{e} +
\cos\theta_{le} = 1$.} 
\end{figure}
For unlike-sign pairs we applied the ``diagonal cut'' $p_e + \cos
\theta_{\ell e} \geq 1$ ($p_e$ in GeV/$c$).  This cut suppressed the
same-$B$ background by a factor of 25, while retaining two-thirds of the
opposite-$B$ unlike-sign electron signal.  The residual contribution of
same-$B$ secondaries that leak through the diagonal cut is small and is
estimated with Monte Carlo normalized to the data as described in
Sec.~\ref{sec:spectrum_corrected}. 
We performed extensive Monte Carlo
studies of potential bias that might have been introduced into our
analysis by this cut.  Semileptonic decays $B \rightarrow X_c \ell \nu$
in $B {\bar B}$ events were simulated as a mixture of resonant and
nonresonant decays.  These used HQET and the
CLEO-measured form-factor 
parameters for $B \rightarrow D \ell \nu$~\cite{Athanas:1997eg} and $B
\rightarrow D^* 
\ell \nu$~\cite{Duboscq:1996mv}, and models for $B \rightarrow D^{**} \ell
\nu$~\cite{Scora:1995ty} and nonresonant modes $B \rightarrow D X \ell
\nu$~\cite{Goity:1995xn}.  These studies 
demonstrated that the efficiency was essentially independent of the
$B$-decay mode.  Different backgrounds were affected quite differently
by this cut, however, and these effects were included in the associated
systematic uncertainties.  This is discussed in
Sec.~\ref{sec:spectrum_systematics}.  

Because the diagonal cut largely eliminated the same-$B$ background from
the unlike-charge sample, the electron spectra for events with unlike-sign
tags ($\frac{dN(\ell^{\pm}e^{\mp})}{dp}$) and for events with like-sign tags
($\frac{dN(\ell^{\pm}e^{\pm})}{dp}$) included 
only primary $B$ semileptonic decays and secondary charm semileptonic
decays from events in which the tag lepton and the signal electron were
daughters of different $B$ mesons.  Assuming universality of the
secondary-charm lepton spectra (we discuss the validity of this
assumption below), Eqs.~(\ref{eq:unlike_1})
and (\ref{eq:like_1}) 
provide the connection between these measured spectra and the differential
branching fractions for primary ($\frac{d\mathcal{B}(b)}{dp}$)
and secondary ($\frac{d\mathcal{B}(c)}{dp}$) decays.

\begin{equation}
\frac{dN(\ell^{\pm}e^{\mp})}{dp} = 
N_{\ell}\,\eta(p)\,\epsilon(p) \,
\Bigg[{\frac{d\mathcal{B}(b)}{dp}} \, 
(1-\chi) \:+\:
{\frac{d\mathcal{B}(c)^{opp B}}{dp}}\,\chi\Bigg]
\label{eq:unlike_1}
\end{equation}

\begin{equation}
\frac{dN(\ell^{\pm}e^{\pm})}{dp} = 
N_{\ell} \, \eta(p) \,
\Bigg[{\frac{d\mathcal{B}(b)}{dp}}\,\chi+ 
{\frac{d\mathcal{B}(c)^{opp B}}{dp}}\,(1-\chi)\Bigg]
\label{eq:like_1}
\end{equation}

\noindent In these equations, $N_\ell$ is the effective number of tags in the
sample, $p$ is the signal electron momentum, $\eta(p)$ is the efficiency
for reconstructing and identifying 
the electron, $\epsilon(p)$ is the efficiency of the diagonal cut
applied to the unlike-sign sample, and $\chi$ is the $B^0 {\bar B^0}$
mixing parameter multiplied by the fraction of all $B {\bar B}$ events
at the $\Upsilon(4S)$ that are neutral $B$'s.  

We determined $\chi$ by combining several pieces of
experimental information.  The Particle Data Group value for the $B_d^0
\overline{B_d^0}$ mixing parameter is $\chi_d = 0.181 \pm
0.004$~\cite{PDG_2002}.  The charged/neutral $B$ lifetime ratio is
$\tau^\pm/\tau^0 = 1.083 \pm 0.017$~\cite{PDG_2002}.  CLEO has measured the 
ratio of charged to neutral $B$ production at the $\Upsilon(4S)$ to be
${f_{+-} \tau_\pm \over f_{00} \tau_0} = 1.11 \pm
0.08$~\cite{Alexander:2000tb}.  From these inputs we found $\chi = f_{00}
\chi_d = 0.089 \pm 0.004$, which has been used in extracting the primary
and secondary spectra.  

Eqs.~(\ref{eq:unlike_1})
and (\ref{eq:like_1}) were derived under the assumption that
the secondary-charm lepton spectra are the same for charged and neutral 
$B$ events.  This assumption was made for our previous lepton-tagged 
measurement of $B \rightarrow X \ell \nu$~\cite{Barish:1995cx, wang_thesis} 
and is inconsistent with currently available data.

Modifying Eqs.~(\ref{eq:unlike_1}) and (\ref{eq:like_1}) to allow for the
different secondary spectra in charged and neutral events, and solving
the resulting equations for the primary and secondary spectra leads to 
Eqs.~(\ref{eq:primary}) and (\ref{eq:secondary}).

\begin{equation}
\frac{d\mathcal{B}(b)}{dp} = 
\frac{1}{(1 - [\Delta(p) + 1]\,\chi)} 
\frac{1}{N_{\ell}\,\eta(p)} 
\Bigg[
\frac{[1 - \chi\Delta(p)]}{\epsilon(p)} \frac{dN(\ell^{\pm}e^{\mp})}{dp} - \chi\Delta(p)
\frac{dN(\ell^{\pm}e^{\pm})}{dp} 
\Bigg]
\label{eq:primary}
\end{equation}

\begin{equation}
\frac{d\mathcal{B}(c)}{dp} = 
\frac{1}{(1 - [\Delta(p) + 1]\,\chi)} 
\frac{1}{N_{\ell} \, \eta(p)}
\Bigg[
\frac{\chi}{\epsilon(p)} \frac{dN(\ell^{\pm}e^{\mp})}{dp} - 
(1 - \chi) \frac{dN(\ell^{\pm}e^{\pm})}{dp} \Bigg]
\label{eq:secondary}
\end{equation}

The new factor $\Delta(p)$ accounts for the secondary-spectra differences 
in charged and neutral events.  We determined $\Delta(p)$ with Monte Carlo
simulations incorporating all relevant information on charm and $B$ production
and decay at the $\Upsilon(4S)$ as compiled by the Particle Data
Group~\cite{PDG_2002}. 
Specifically, $\Delta(p)$ reflects the combined effect of the different
branching fractions for $B^0 \to \bar{D^0} X$, $B^0 \to D^- X$, $B^+ \to
\bar{D^0} X$, and $B^+ \to D^- X$, the difference between the 
semileptonic branching fractions of charged and neutral $D$'s, and $B^0 {\bar B^0}$ 
mixing.  Fig.~\ref{fg:Delta} shows the $\Delta(p)$ obtained in our study.
\begin{figure}[ht]
\begin{center}
\epsfysize=10 cm
\epsffile{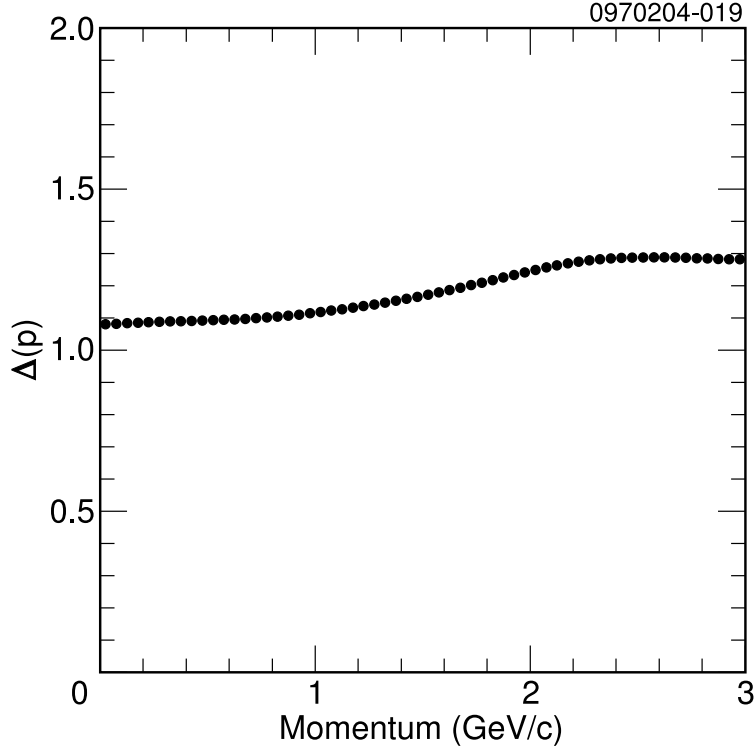}
\caption[fg:Delta]
{\label{fg:Delta} Secondary correction factor $\Delta(p)$. } 
\end{center}
\end{figure}
The systematic uncertainty introduced by this correction was assessed as
half of the difference between results obtained with $\Delta(p)$ as
shown in Fig.~\ref{fg:Delta} and those obtained with $\Delta(p)=1$,
which recovers the previous assumption.  
More detail on this correction can be found in Ref.~\cite{stepaniak_thesis}.  

In the following three sections we describe the determination of the
charge-separated spectra, their backgrounds, the efficiencies, and the final
extraction of the primary spectrum.  The systematic uncertainties that affect all 
quantities derived from the measured primary spectrum are discussed in 
Sec.~\ref{sec:spectrum_systematics}.  

\subsection{Charge-Separated Spectra and Background Corrections}
\label{sec:spectrum_corrected}
The raw $\Upsilon(4S)$ electron momentum spectra for the unlike-sign
sample with the diagonal cut applied and for the like-sign sample are
shown in Fig.~\ref{fg:raw}.  
\begin{figure}[hbtp]
\centerline{
\begin{minipage}{9 cm}
\epsfxsize=9 cm
\epsfysize=9 cm
\epsffile{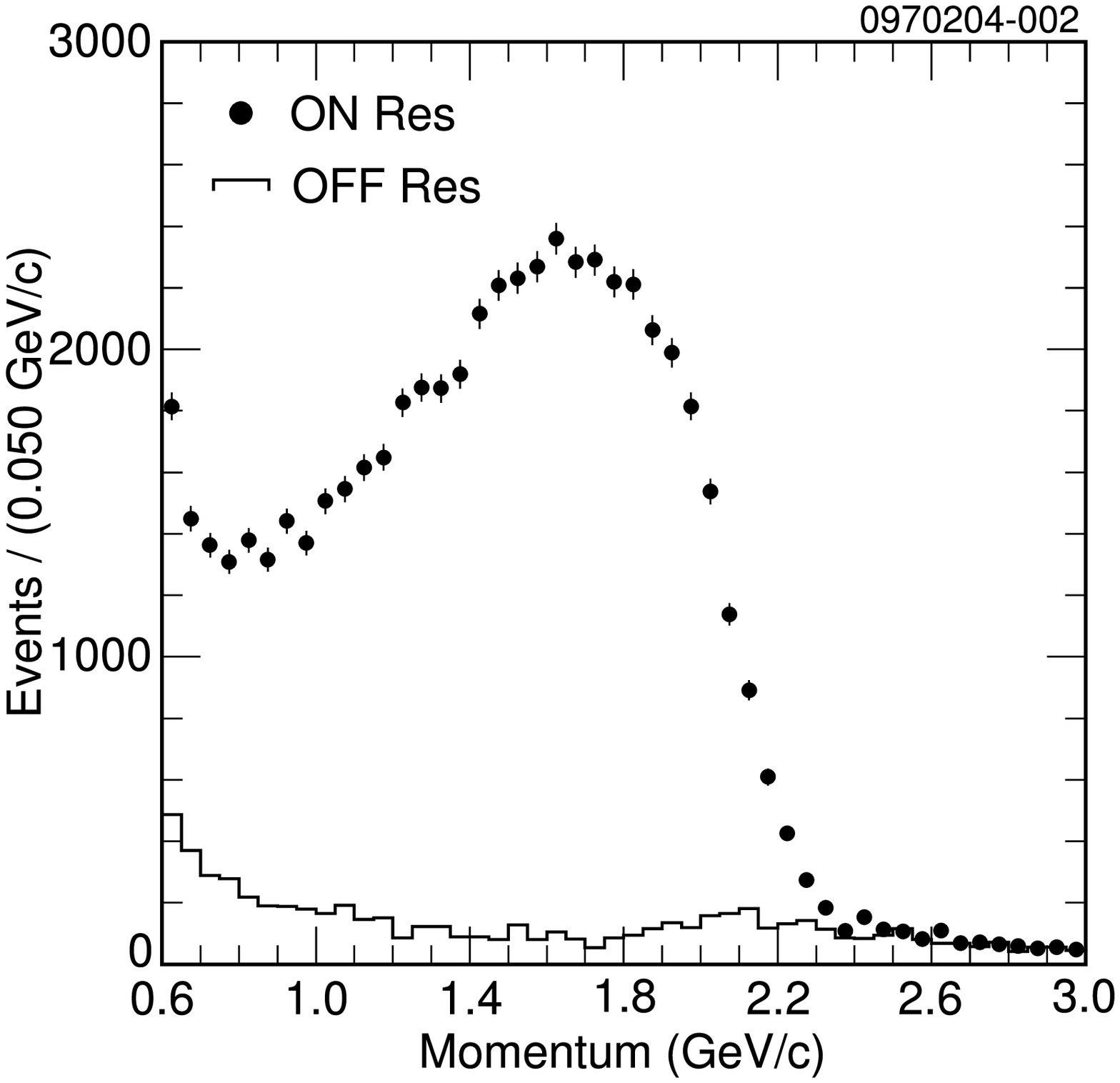}
\end{minipage}
\begin{minipage}{9 cm}
\epsfxsize=9 cm
\epsfysize=9 cm
\epsffile{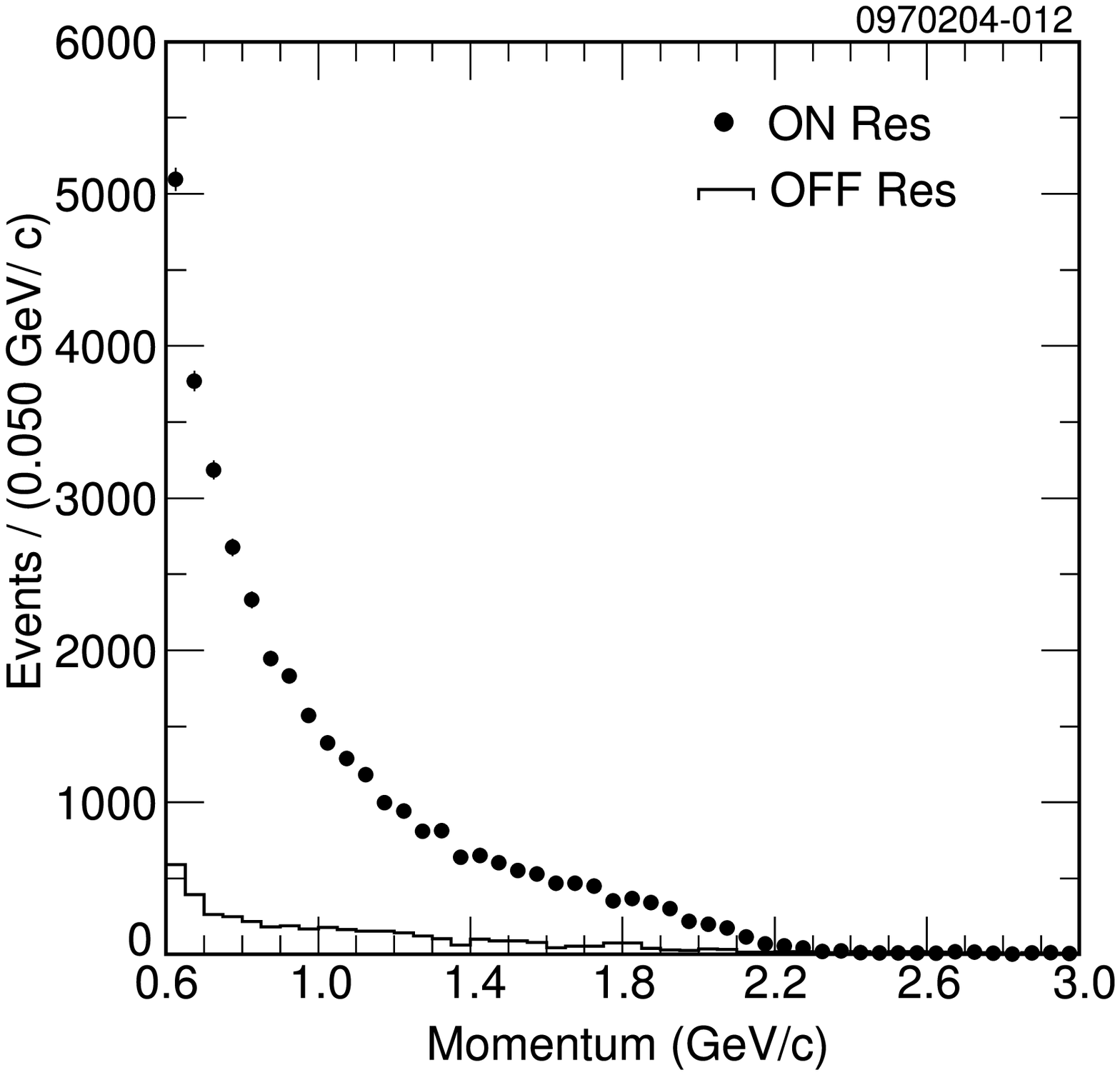}
\end{minipage}
} 
\caption[fg:raw] {\label{fg:raw} Electron-momentum spectra for (left)
unlike-sign pairs 
passing the  diagonal cut, and (right) like-sign pairs without the cut.
The points represent data collected on the $\Upsilon(4S)$ peak and the
histograms are the estimated continuum contributions determined with
scaled below-resonance data.} 
\end{figure}
These raw spectra include several backgrounds that had to be subtracted
before the $B \rightarrow X e \nu$ spectrum could be obtained.  Some of
this background was due to real electrons that entered the sample
because of false muon or electron tags.  The false tags included hadrons
misidentified as leptons (``fakes'') and real leptons from processes
other than semileptonic $B$ decays.  Among the latter were leptons from
semileptonic decays of charmed particles, leptons from $J/\psi$ decays,
$\pi^0$ Dalitz decays and photon conversions that leaked through one of
the vetoes, and leptons from other sources in $B$ decays, including
leptonic decays of $\tau$, leptonic decays of $\psi'$ and Dalitz decays
of $\eta$.  The minimum momentum requirement for tag selection of 1.4~GeV/$c$
ensured that these backgrounds were small.  

Background processes contributing directly to the signal
electrons for events with true lepton tags were somewhat larger.  These
included fakes, the sources of real leptons listed above as contributing
to the tags, and several other mechanisms yielding real electrons.  Most
charmed-meson semileptonic decays were not treated as background, but
were isolated algebraically using Eqs.~(\ref{eq:primary})
and~(\ref{eq:secondary}) as described in
Sec.~\ref{sec:spectrum_extracted}.  Three sources of electrons from charm 
were subtracted as backgrounds.  The first was the small component of 
unlike-sign electrons from same-$B$ charm decays that passed the diagonal cut.  
The second was electrons from decays of ``upper-vertex'' charm daughters of 
the other $B$ ($b \rightarrow c W^+$, $W^+ \rightarrow c {\bar s}$),
which was an unlike-sign contribution that could not be distinguished 
kinematically from the $B \rightarrow X e \nu$ signal.  The third was 
electrons from the decay of charmed-baryons.  

The background due to both tag and signal fakes in the $B {\bar B}$
spectra was estimated by 
combining misidentification probabilities per track, binned in momentum,
with the hadronic track momentum spectra obtained from data by imposing all
selection criteria except for lepton identification.  These track
spectra were corrected for the contributions of real leptons.  The
misidentification probabilities were measured with samples of
pions from reconstructed $K^0_S$ decays, kaons from $D^* \rightarrow D
\rightarrow K \pi$ and protons and antiprotons from the decays of
$\Lambda$ and ${\bar \Lambda}$.  Monte Carlo simulations were used to correct the
measured muon misidentification probabilities for the small
underestimate that resulted when pion or kaon decays in flight prevented
the successful reconstruction of the $K^0_S$ or $D$, but not the
misidentification as a muon.  Relative particle abundances as a
function of momentum were determined
with Monte Carlo and used to combine the measured pion, kaon and $p/{\bar p}$
fake rates into misidentification probabilities per hadron track that
were appropriate for $B$ decays.  

The backgrounds due to veto leakage in the tag and signal samples were
estimated by Monte Carlo 
simulation.  The normalization for this correction was determined from
data by fitting the spectra of vetoed leptons in Monte Carlo to the
corresponding spectra in the data.  The fits demonstrated that the Monte
Carlo does a very good job of reproducing the observed distributions, in 
particular for $J/\psi$, which is the most important veto.  

The leakage of same-$B$ secondary signal electrons was estimated with a
procedure similar to that for the veto leakage.  In this case, the
two-dimensional distribution of $\cos\theta_{\ell e}$ versus
signal-electron momentum was fitted.  Again, the normalization was
determined by fitting the Monte Carlo distributions for same-$B$
secondary signal electrons that failed the diagonal cut to the
corresponding distribution in data.  This factor was then used to scale
the Monte Carlo distributions for those that leaked through the cut,
providing the background correction that was applied to the electron
spectrum.  

Other physics backgrounds to both tags and signals were estimated with Monte
Carlo simulations, primarily a sample of ``generic'' $B {\bar B}$ events
with neutral $B$ mixing modeled to agree with present experimental
observations.  This simulated sample had 
five times the statistics of $\Upsilon(4S)$ data sample.  

Fig.~\ref{fg:bgs} shows the continuum-subtracted unlike-sign and 
\begin{figure}[hbtp]
\centerline{
\begin{minipage}{9 cm}
\epsfxsize=9 cm
\epsfysize=9 cm
\epsffile{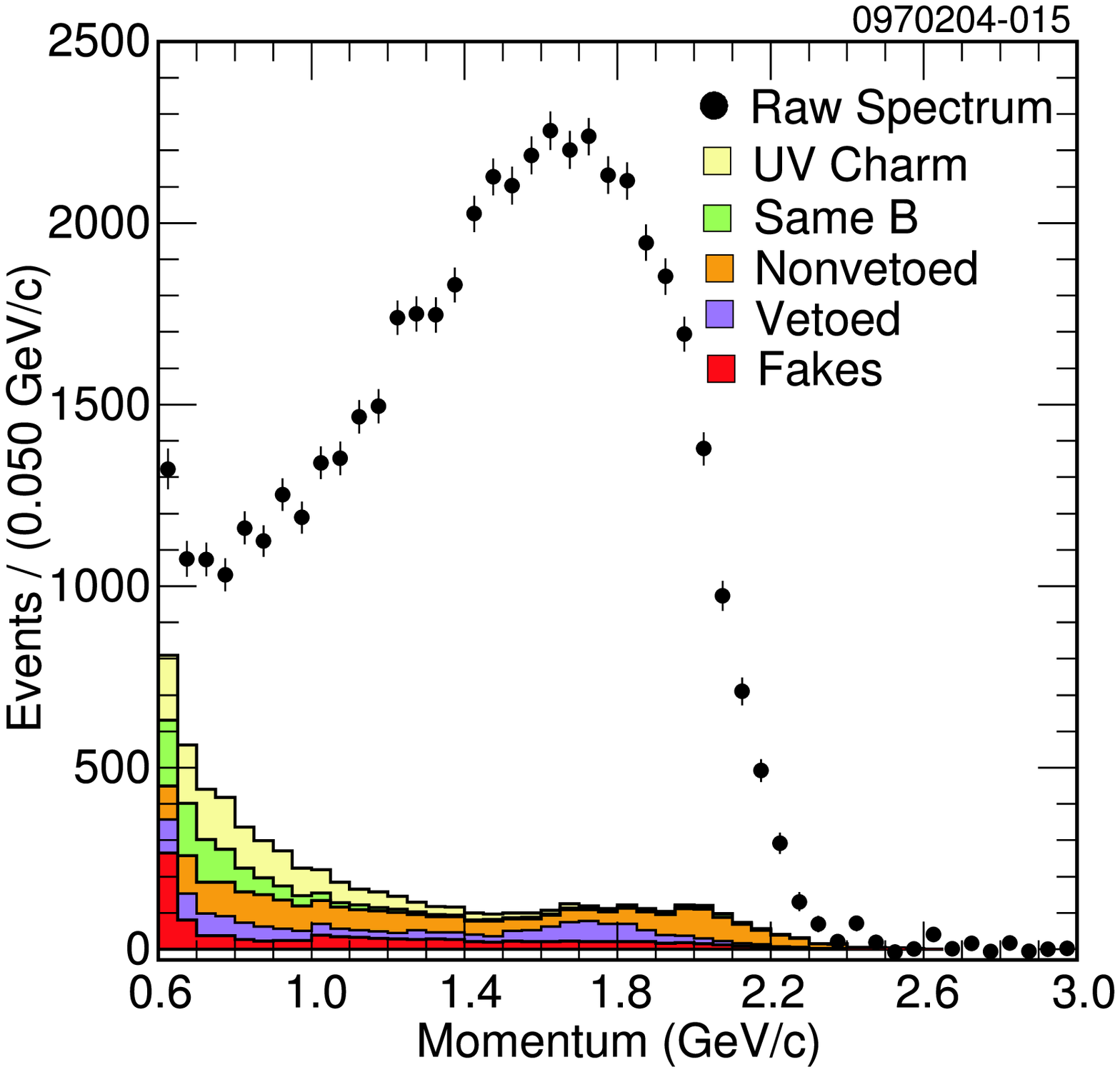}
\end{minipage}
\begin{minipage}{9 cm}
\epsfxsize=9 cm
\epsfysize=9 cm
\epsffile{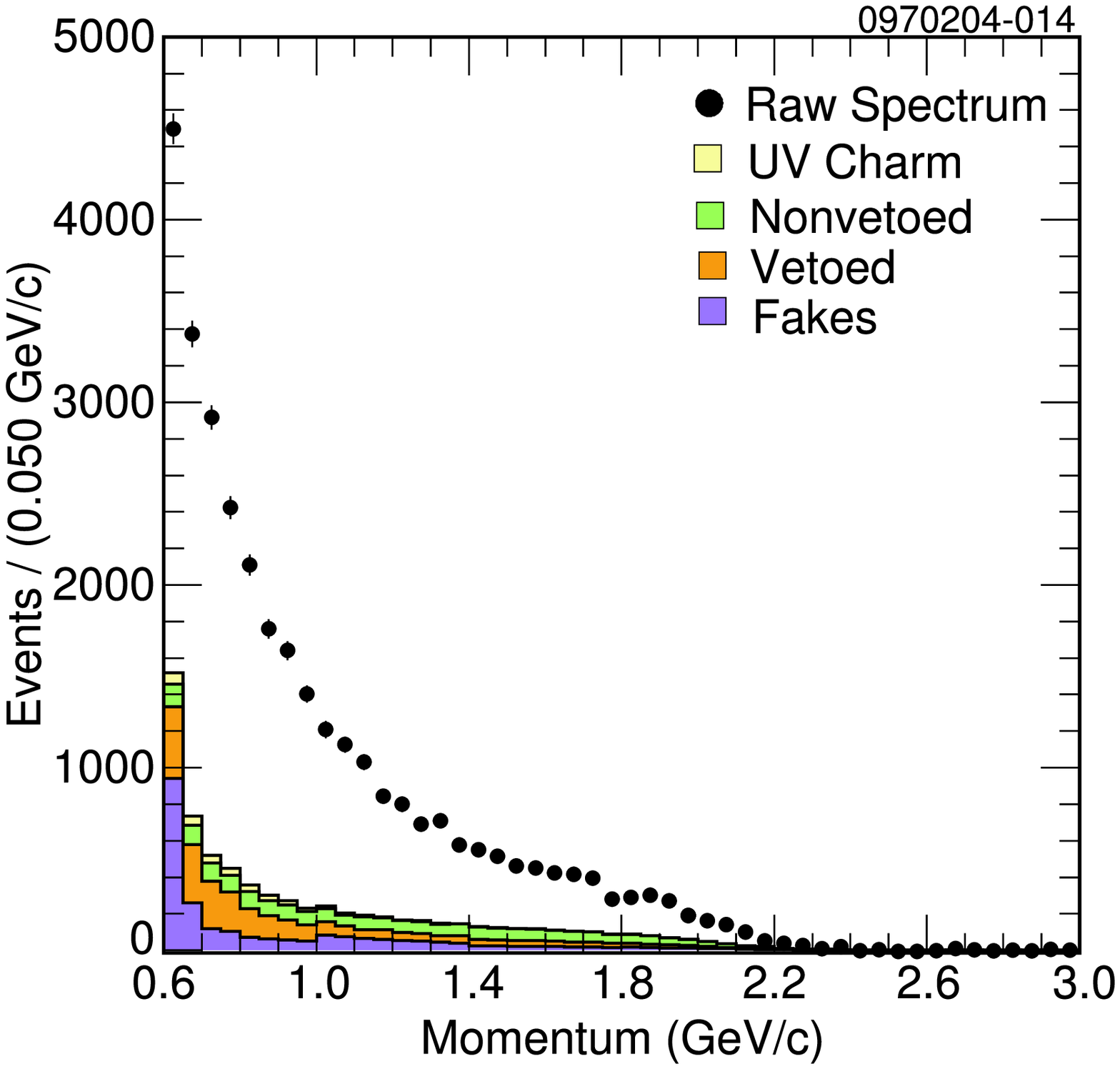}
\end{minipage}
} 
\caption[fg:bgs]
{\label{fg:bgs} Continuum-subtracted unlike-sign (left) and like-sign
(right) spectra, showing the breakdown of backgrounds computed as
described in the text.} 
\end{figure}
like-sign spectra together with the backgrounds determined with the
procedures described above.  Sources of both tag-lepton and
signal-electron backgrounds have been combined in these plots.  For
example, 
electrons that are the direct product of an upper-vertex charm decay and
electrons that are accompanied by a tag from an upper-vertex
charm decay are both included in the category ``UV charm.''
The spectra after the subtraction of all backgrounds are shown in
Fig.~\ref{fg:bg_subtracted}.  
\begin{figure}[hbtp]
\centerline{
\begin{minipage}{9 cm}
\epsfxsize=8.5 cm
\epsfysize=8.5 cm
\epsffile{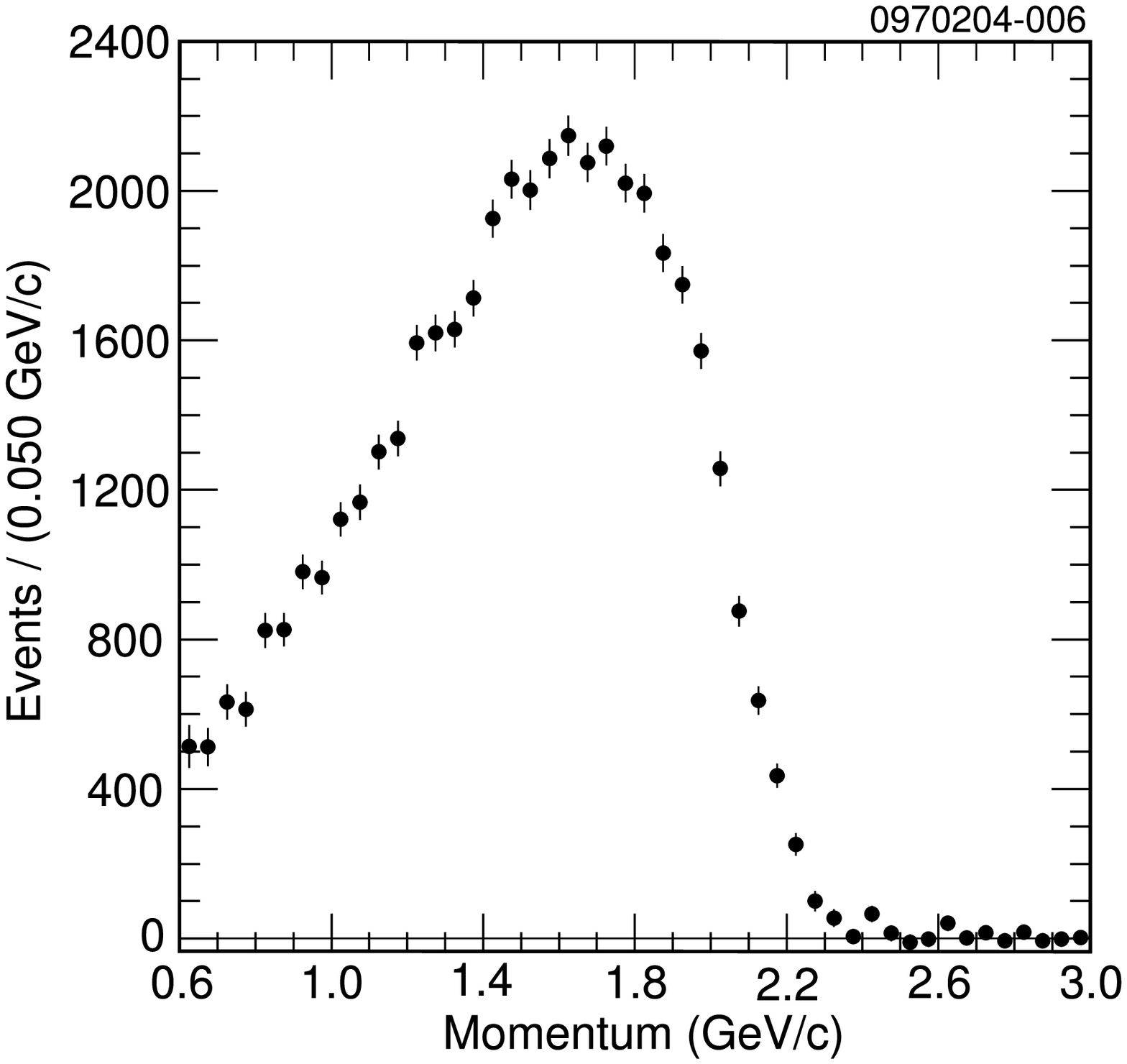}
\end{minipage}
\begin{minipage}{8.5 cm}
\epsfxsize=8.5 cm
\epsfysize=8.5 cm
\epsffile{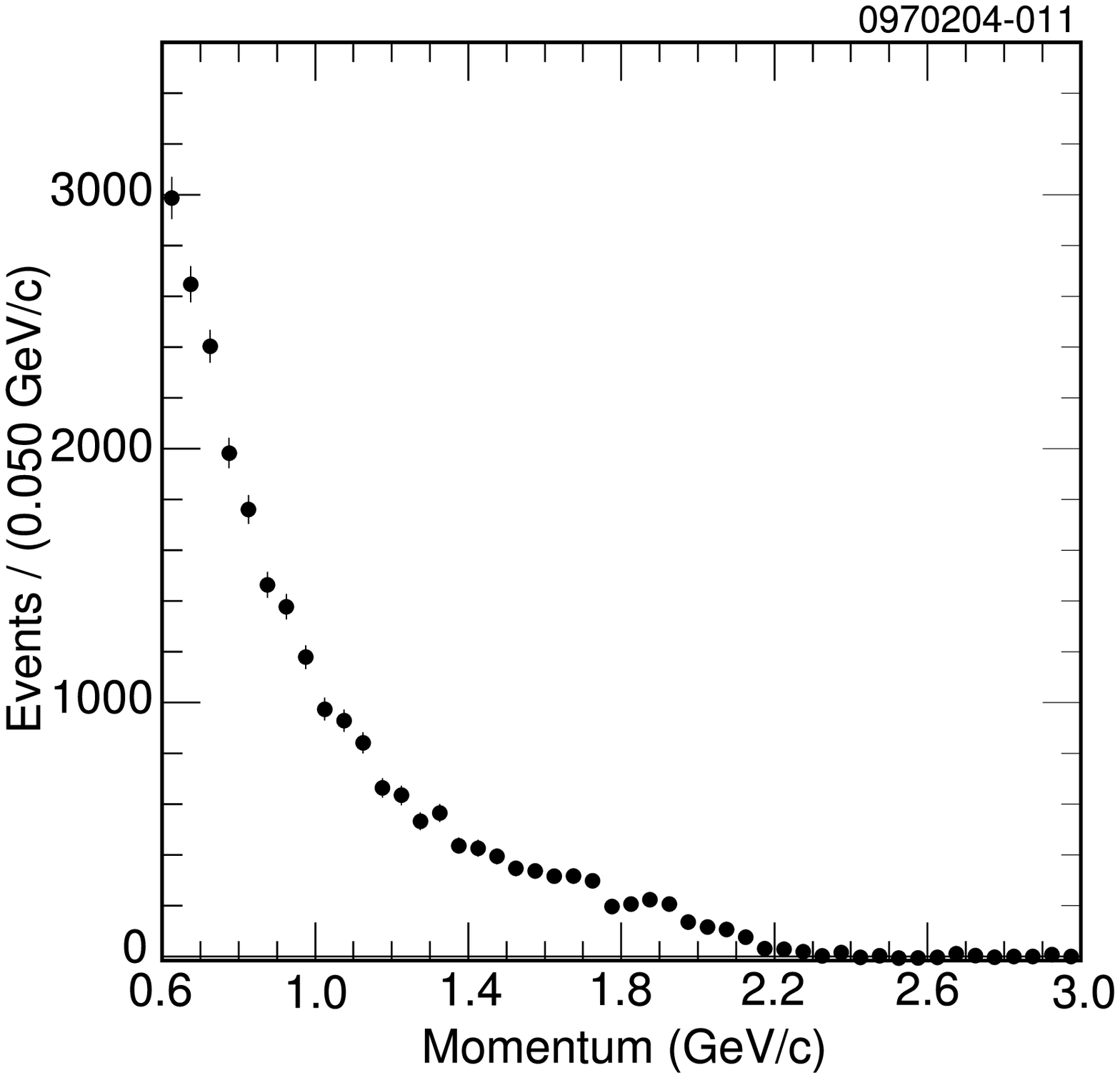}
\end{minipage}
} 
\caption[fg:bg_subtracted]
{\label{fg:bg_subtracted} Unlike-sign (left) and like-sign (right)
electron spectra after all backgrounds have been subtracted.  These are 
the spectra that were passed to Eqs.~(\ref{eq:primary}) and  
(\ref{eq:secondary}).} 
\end{figure}
Systematic uncertainties in the background corrections are described in
Sec.~\ref{sec:spectrum_systematics}.   
 
\subsection{Counting Tags}
\label{sec:spectrum_counting_tags}
The normalization for the measurement of the $B$ semileptonic branching
fraction is provided by $N_{\ell}$, the effective number of tags in our
lepton-tagged event sample.  Its determination is given in Table~\ref{fg:tag_count_table}.
Identified leptons satisfying the tag requirements of
Sec.~\ref{sec:event_selection} were counted for both the
on-$\Upsilon(4S)$ and below-resonance data samples.  After correction
for the continuum, fake leptons, and other backgrounds by the procedures
described in Sec.~\ref{sec:spectrum_corrected}, the raw number of tags
from semileptonic $B$ decays was found to be $N_{\ell}^{\rm raw} =
1,137,042 \pm 1631$, where the error is statistical only.

\begin{table}[hbtp]
\begin{center}
\begin{tabular}{|l|c|c|c|}
\hline
Source & $\mu$ & $e$ & $\mu + e$\\ \hline
ON $\Upsilon(4S)$   & 828,155 $\pm$ 910   & 837,002 $\pm$ 915   & 1,665,157 $\pm$ 1,290\\ \hline
Scaled Continuum    & 261,667 $\pm$ 737   & 212,146 $\pm$ 664   & 473,813 $\pm$ 992 \\ \hline
Cont. Subtracted    & 566,488 $\pm$ 1,171 & 624,856 $\pm$ 1,131 & 1,191,344 $\pm$ 1,628 \\ \hline \hline

Fake Leptons        & 11,385 $\pm$ 61    & 936 $\pm$ 4    & 12,321 $\pm$ 61 \\ \hline
$J/\psi$            & 3,397 $\pm$ 28     & 4,451 $\pm$ 31 & 7,848 $\pm$ 42   \\ \hline
$\pi^0$             & 	    N/A          & 190 $\pm$ 8    & 190 $\pm$ 8  \\ \hline
$\gamma$            & 	    N/A          & 116 $\pm$ 6    & 116 $\pm$ 6  \\ \hline

Secondary Charm     & 10,484 $\pm$ 47    & 13,347 $\pm$ 52 & 23,831 $\pm$ 70\\ \hline
Upper-Vertex $D$    & 330 $\pm$ 9        & 417 $\pm$ 9     & 747 $\pm$ 13 \\ \hline
Upper-Vertex $D_s$  & 2,364 $\pm$ 22     & 818 $\pm$ 13    &  3,182 $\pm$ 26 \\ \hline
$\tau$              & 1,947 $\pm$ 20     & 2,538 $\pm$ 22  & 4,485 $\pm$ 30 \\ \hline
$\psi'$             & 588 $\pm$ 11       & 609 $\pm$ 11    & 1,197 $\pm$ 16 \\ \hline
Other Backgrounds   & 356 $\pm$ 9        & 29 $\pm$ 3      & 385 $\pm$ 9 \\ \hline
\hline
Background-Subtracted Yield     & 535,637 $\pm$ 1,174 & 601,405 $\pm$ 1,132 & 1,137,042 $\pm$ 1,631\\ \hline 
\end{tabular}
\caption{\label{fg:tag_count_table}{Yields and backgrounds for tag count.
  Errors are statistical only.}}
\end{center}
\end{table}
It was not necessary to correct the tag count for the absolute 
efficiencies of lepton selection, such as track-quality requirements
and lepton identification, because the background-corrected sample of
events with tags provides us with $B {\bar B}$ events in
which one $B$ is known to have decayed semileptonically.  It is the
fraction of these events in which the other $B$ decayed to an electron
that gives the semileptonic branching fraction.  The only necessary
corrections to the tag count are for effects that result preferentially
in the gain or loss of events in which both $B$'s decayed
semileptonically.  

Such a correction to the tag count was necessitated by the effect of the
charged multiplicity requirement in the event selection, since
semileptonic decays typically have lower multiplicity than hadronic
decays.  We evaluated this effect with a large sample of simulated $B
\bar{B}$ events.  The event-selection efficiency $\epsilon_{\ell}$ for
any event with a 
lepton tag from semileptonic $B$ decay was found to be 95.8\%, while
the efficiency $\epsilon_{\ell e}$ for events with a lepton tag and a
second semileptonic $B$ decay was 91.0\%.  This gives a relative
event-selection efficiency of $\epsilon_r = \epsilon_{\ell
  e}/\epsilon_{\ell} = 95.0\%$, showing that our direct tag count
was an overestimate of the true number of events with tags that could
enter our primary spectrum. Therefore, the effective number of tags was
$N_{\ell} = \epsilon_r N_{\ell}^{\rm raw} = 1,079,901 \pm 1,549$
(statistical uncertainty only).  

This relative event-selection efficiency introduced a systematic
uncertainty into our measurement associated with how well the
Monte Carlo simulated the multiplicity of both hadronic and semileptonic
$B$ decays.  We compared the observed
charged multiplicity distributions for $B {\bar B}$ events in 
data and in Monte Carlo and found the agreement to be
quite good.  The measured mean multiplicities agreed within 0.1 unit for 
all events with tags, and within 0.01 unit for events with tags and
electrons from $B \rightarrow X e \nu$.  The latter difference was
determined to be negligible, and the systematic uncertainty associated
with the former was assessed by reweighting the Monte Carlo sample in
event multiplicity.  

We note here that there was a misconception in the treatment of this
effect in our previous analysis~\cite{Barish:1995cx}, which is
superseded by this paper.  In that case, the relative
event-selection efficiency was calculated with a numerator that included
all signal electrons, not just the primary $B \rightarrow X e \nu$
electrons.  Including all dilepton events in the numerator had the
effect of raising the average charged multiplicity in those events,
since it admitted cases where an electron is produced further down the
decay chain, with more accompanying hadrons.  When calculated in this
incorrect way, the relative event-selection efficiency was overestimated
and the semileptonic branching fraction underestimated by a few percent
relative. 

\subsection{Efficiencies and Extracted Primary and Secondary Spectra}
\label{sec:spectrum_extracted}
To extract the primary and secondary spectra, the remaining step was the
substitution of our corrected yields into Eqs.~(\ref{eq:primary}) and
(\ref{eq:secondary}). 
In addition to the quantities already given, this required determination of
the efficiencies $\eta(p)$ and $\epsilon(p)$ for the detection of the electron 
and the effect of the diagonal cut on the opposite-sign sample, respectively.
The electron detection efficiency $\eta(p)$ includes the efficiency of
the fiducial cut on electron candidates, 
the efficiency of track-quality cuts, the efficiency of the electron identification,
and the efficiency for passing the three vetoes ($J/\psi$, $\pi^0$ Dalitz, 
$\gamma$-conversion).  Each of these, except for the electron identification, was
obtained by processing Monte Carlo simulations of $\Upsilon(4S)$ events.  Where 
possible, the Monte Carlo was normalized or validated with data.  The bin-by-bin
effect of bremsstrahlung in the detector material was also incorporated into the
efficiency through this simulation.

Studies of electron-identification and track-selection efficiencies were
performed with tracks from radiative Bhabha events embedded into
hadronic events.  The ``target'' hadronic events were selected to ensure
that the final embedded samples were compatible with $B {\bar B}$ signal
events in event topology, multiplicity and electron angular
distribution.  For the tracking studies, embedded samples were prepared
for both data and Monte Carlo, and comparison of the two gave a
correction factor as a function of electron momentum that could
subsequently be applied to the efficiency determined with simulated
signal events.  For the track-selection criteria used in this analysis,
the correction factor proved to be almost negligibly different from unity. 
 
The embedded radiative Bhabha sample was also used to measure the 
efficiency of our electron-identification package.  In this case the
efficiency determined for electrons in the embedded sample was applied
directly to data, and extensive studies were made of systematic
uncertainties.  These studies are described in
Sec.~\ref{sec:spectrum_systematics}. 


With all ingredients assembled, the final step was substitution into 
Eqs.~(\ref{eq:primary}) and (\ref{eq:secondary}) to obtain the separated primary 
and secondary spectra.  These are shown in Fig.~\ref{fg:solved}.  The
apparent pairing of points on the rising side of the primary spectrum
has been studied extensively.  It is not attributable to any one step of
the analysis procedure, and we have found no other explanation other
than a statistical fluctuation.
\begin{figure}[htp]
\centerline{
\begin{minipage}{8.5 cm}
\epsfxsize=8.5 cm
\epsfysize=8.5 cm
\epsffile{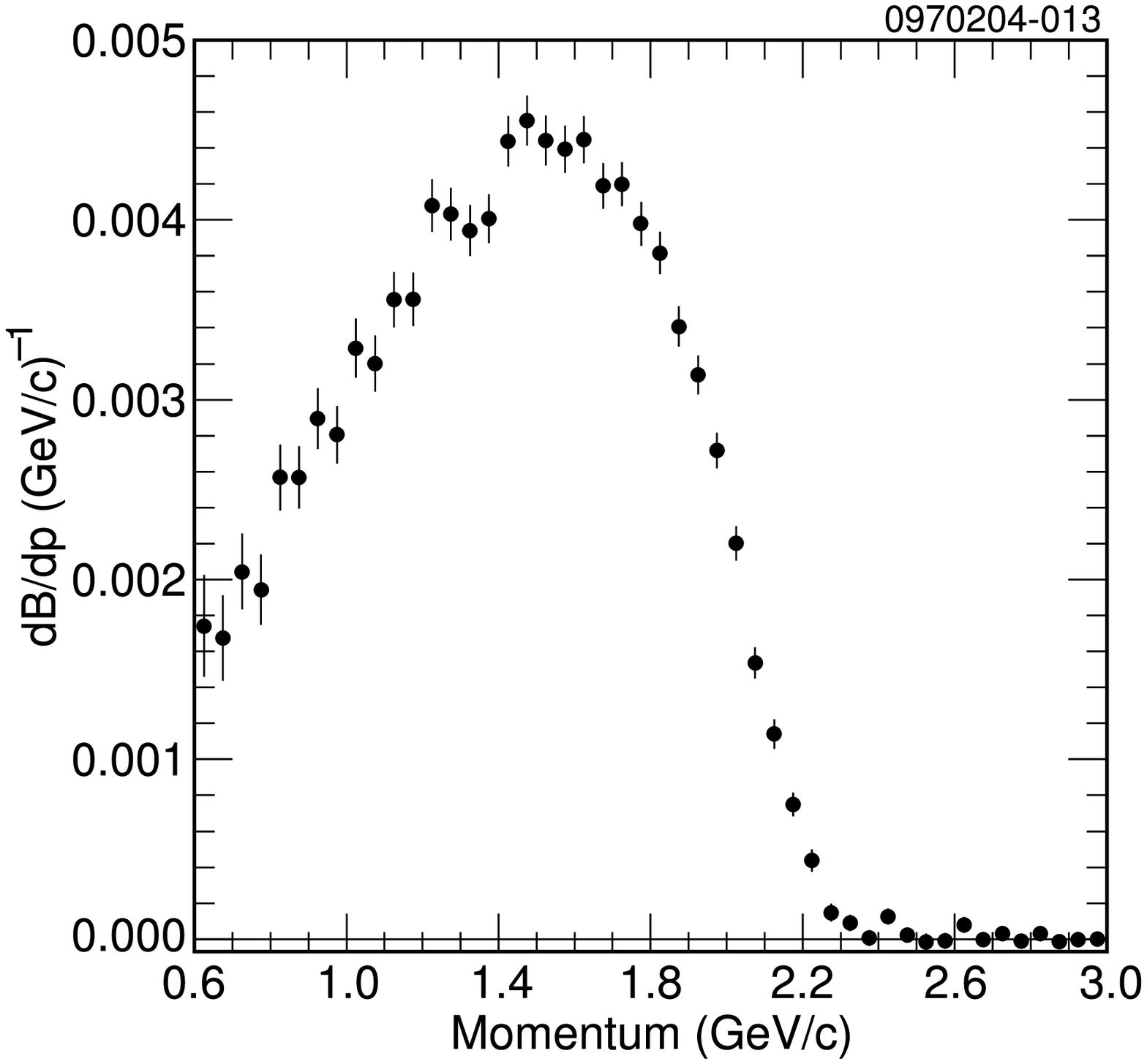}
\end{minipage}
\begin{minipage}{8.5 cm}
\epsfxsize=8.5 cm
\epsfysize=8.5 cm
\epsffile{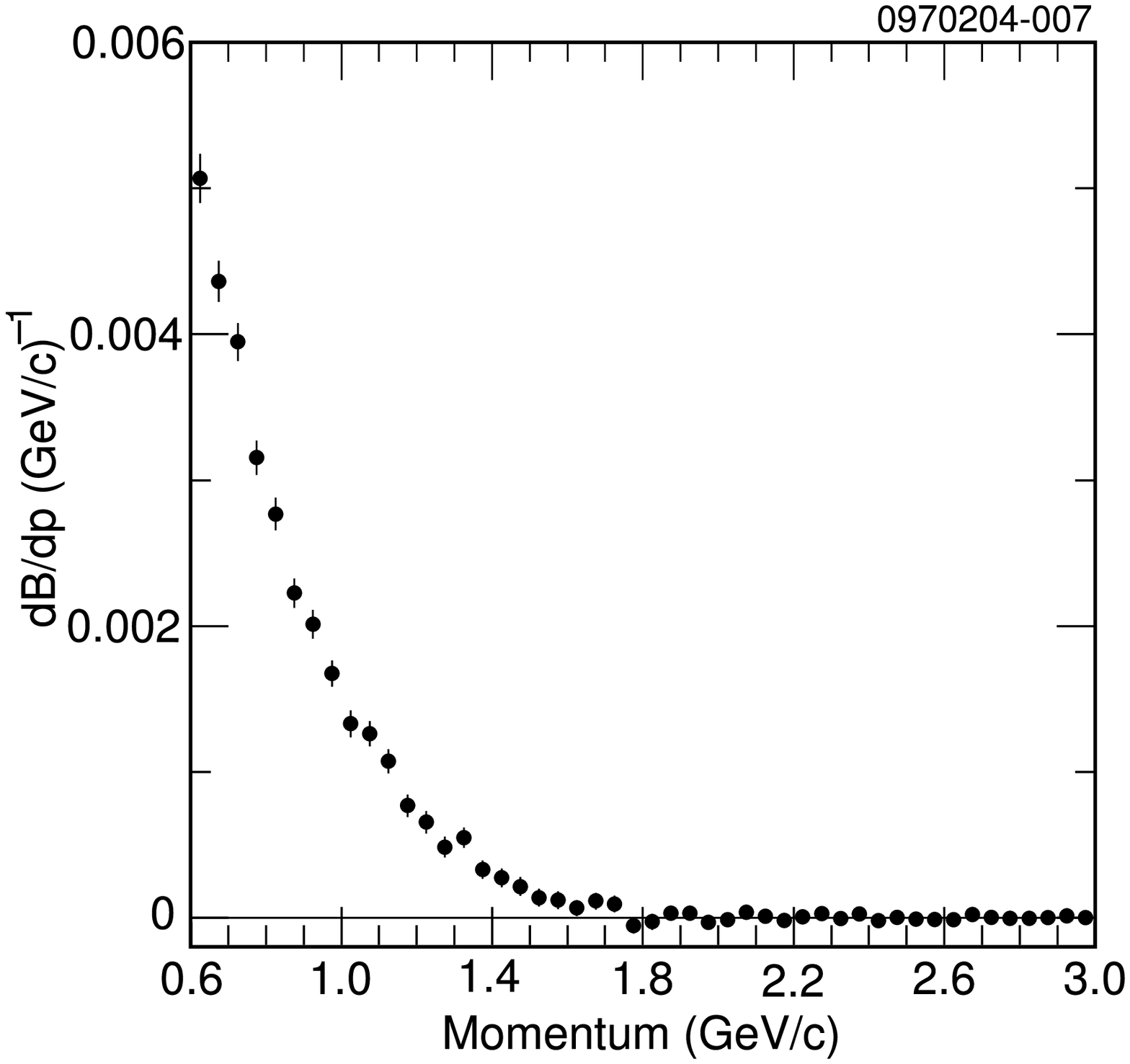}
\end{minipage}
}
\caption[fg:solved]
{\label{fg:solved} Primary (left) and secondary (right) spectra,
  obtained by solving Eqs.~(\ref{eq:primary}) and (\ref{eq:secondary}).}
\end{figure}
Sec.~\ref{sec:SLBR} and Sec.~\ref{sec:moments} describe the extraction of the 
$B \rightarrow X e \nu$ branching ratio and the electron-energy moments from the primary 
spectrum, respectively.  Sec.~\ref{sec:spectrum_systematics} provides
details on the systematic
uncertainties of the spectrum measurement that are common to both.

\section{Systematic Uncertainties and Cross-checks}
\label{sec:spectrum_systematics}
Nearly all of the systematic uncertainties in the measurements of the $B$
semileptonic branching fraction and the electron-energy moments are rooted
in the systematic uncertainties in the spectrum measurement.  Many of these
have already been identified, and this section provides additional details
about their evaluation.  The actual systematic uncertainty estimates are 
presented in Sec.~\ref{sec:SLBR} and Sec.~\ref{sec:moments}.  Full
details of the systematic  
studies are available in Ref.~\cite{stepaniak_thesis}.  

\subsection{Veto-Leakage Corrections}
These corrections were computed using momentum spectra determined from Monte
Carlo simulations with normalizations obtained by fitting data, as described in 
Sec.~\ref{sec:spectrum_corrected}.  This procedure ensured that the 
corrections  were insensitive to uncertainty in the rates of the contributing 
processes, although there remained some sensitivity to the modeling of
details like the momentum spectra.  The $J/\psi$ modeling is believed to
be very accurate: the mixture of decays was tuned to agree with exclusive 
branching ratios~\cite{PDG_2002} and the inclusive $J/\psi$ momentum
spectrum~\cite{Anderson:2002jf}.  We estimated 
a $\pm 5\%$ systematic uncertainty on the subtraction of unvetoed 
$J/\psi$'s.  For the $\pi^0$ and photon-conversion vetoes,
there was more uncertainty in the simulation of the detector response,
and we took $\pm 20\%$.  For each of these, we have fluctuated the correction
upward and downward by these amounts and taken the systematic uncertainty
on any observable to be one-half of the difference between them.

\subsection{Same-$B$ Secondaries}
\label{sec:diag_cut_leakage_syst}
The background due to same-$B$ secondaries that were not eliminated by 
the diagonal cut was also computed with Monte Carlo normalized to data,
as described in Sec.~\ref{sec:spectrum_corrected}.
In this case, the yield and distribution for the same-B secondaries that
were successfully cut (98\%) were used to normalize the distribution for
those that leaked through (2\%), with negligible statistical
uncertainty.  An excellent fit was obtained in the two 
dimensions of opening angle versus momentum, demonstrating that the 
Monte Carlo did a very good job of reproducing the detailed distributions 
of the contributing processes.  The systematic uncertainty for this
correction was taken to be 15\%.

\subsection{Other Non-Vetoed Background Corrections}
Similar to the method of determining the systematic errors attached to
veto leakage, we used the Monte Carlo to simulate the shapes of the momentum
spectra for backgrounds due to non-vetoed physics processes.  For each component 
we attempted to assess a reasonable uncertainty based on world-average branching 
fractions and other information.  In all cases we take as the systematic uncertainty 
one-half of the difference between the extreme variations.

Upper-vertex charm was the largest of these sources.  Broadly speaking,
this background can be broken down into two components: final states 
with a $D_s$ meson and another charmed particle and final states with
two non-strange charmed mesons.  We treated these independently,
since their estimates are largely based on different experimental and 
theoretical inputs.  While the semileptonic branching fraction 
${\mathcal B}(D_s \to X e \nu$) is not well measured, the $D^0$ and $D^+$ 
semileptonic branching fractions can be combined with lifetime data to 
estimate ${\mathcal B}(D_s \to X e \nu) \simeq 8\%$, an estimate that is 
probably reliable at the 10\% level.  However, this uncertainty is essentially 
negligible compared to that in the branching fraction for 
$B \to D_s X$, which has been estimated to be 
$9.8 \pm 3.7$\%~\cite{LEP_WG}, based on a variety of exclusive measurements.  
Using these assumptions, we took the overall systematic 
uncertainty on the contribution of semileptonic decays of upper-vertex 
$D_s$ to be $\pm 40\%$.

The upper-vertex $D$ contribution is somewhat better known, with
well-measured semileptonic branching fractions~\cite{PDG_2002} and an
estimated rate for $B \to {\bar D}D^{(*)}X$ of $8.2 \pm 1.3$\%
\cite{LEP_WG}.  We assigned a systematic  uncertainty to the electrons
from upper-vertex non-strange charmed mesons of $\pm 25\%$.

The estimated contributions of $B \to \tau \to e$ and 
$B \to \psi' \to e^+e^-$ were both based on world-average measured
branching fractions~\cite{PDG_2002}.  Both were assigned systematic errors of $\pm 15\%$, 
taking into account the errors of those branching fractions, with some
additional uncertainty associated with the shapes of the momentum spectra. 

\subsection{Lepton Identification}
\label{sec:sys_fakes}
Since muons were only used for tags, the correction for fake muons only
entered our results through the normalization of the primary spectrum.
We took an overall systematic uncertainty in the estimate of muon fakes of $\pm
25\%$.   The muon-identification efficiency was not used in our measurement.

For our previous lepton-tagged analysis \cite{Barish:1995cx}, the results
obtained were yields and branching fractions with sensitivity only to
the momentum-averaged efficiency.  It was therefore unnecessary to
scrutinize carefully the 
reliability of the measured momentum dependence of the  
electron-identification efficiency. The determination of the 
spectral moments of the electron energy spectrum was much more demanding 
in this regard.  As has been described in Sec.~\ref{sec:event_selection}, 
momentum-dependent biases in the radiative-Bhabha-measured efficiency for
the standard CLEO II electron-identification package led us to reoptimize
with simpler criteria.

Two approaches were used to assess the systematic uncertainties in electron
identification.  In the first, estimates were made based on studies of the
radiative Bhabha and tagged-track samples that were used to determine the
efficiency and misidentification probabilities.  These involved techniques
like varying selection cuts and comparison of embedded and unembedded samples
that clearly probed systematic effects, but were difficult to use for a
quantitative assessment.  Overall uncertainties were estimated to be
in the range of 2\% for the electron-identification efficiency.  For
the misidentification probability the uncertainty was estimated to increase
from $25\%$ below 1~GeV/$c$ to 100\% above 1.5~GeV/$c$.  Uncertainty
in the momentum dependence was very difficult to assess.  Monte Carlo
studies were inconclusive, and the effect on the electron-identification
efficiency was bracketed by ``worst-case skewing'' of the radiative 
Bhabha measurement.

This approach was deemed to be unsatisfactory for the moments measurement,
so we developed a second procedure that relied on the ``factorizability''
of our simplified electron identification.  Each of the component criteria 
of the electron identification ($dE/dx$ requirement, 
low-side $E/p$ cut, high-side $E/p$ cut, time-of-flight, likelihood cut
for momenta below 1~GeV/$c$), was separately adjusted and the entire analysis, 
including efficiency and fake-rate determinations, was repeated.  The amount 
of ``knob-turning'' was determined based on the inefficiency associated with 
each cut, which was typically a few per cent.  The target was a 
tightening of the cut sufficient to double its inefficiency.  In the cases 
of the less powerful elements of the selection ($dE/dx$ and time-of-flight), 
the alternative was to turn off that cut completely.  The
resulting primary spectra were processed to obtain the observables of
our analysis, the branching fraction and moments, and the difference 
between the results for the standard and modified analyses was taken
as the systematic uncertainty associated with that component of the electron
identification.  Since the five different knobs represented
independent elements of the electron selection, we combined their
systematic uncertainties in quadrature.

\subsection{Other Efficiency Corrections}
\label{sec:other_eff_syst}

The track-selection efficiency was determined with a Monte Carlo simulation
of signal events, corrected by the data/Monte Carlo ratio determined with 
embedded radiative Bhabha events, as described in
Sec.~\ref{sec:spectrum_extracted}.  
The systematic error associated with this efficiency was assigned to be
the difference between results obtained with the standard spectrum, and
those obtained without application of the data/Monte Carlo correction.
 
We set the systematic uncertainty due to the efficiency of the diagonal 
cut based on extreme variations of the mixture of semileptonic $B$
decays in our simulated event sample.  Variations were constrained
by measured branching fractions \cite{PDG_2002}.  The mixtures considered 
ranged from the ``hardest possible'' primary spectrum 
($B \rightarrow D^* e \nu$ 
increased by 6\%; $B \rightarrow D^{**} e \nu$ increased by 30\%; 
$B \rightarrow D e\nu$ decreased by 8\%; nonresonant 
$B \rightarrow D^{(*)} X e \nu$ decreased by 30\%) to 
the ``softest possible'' primary spectrum (reverse of the above 
variations).  For each case we computed a new diagonal cut efficiency, 
rederived the final spectrum, and calculated new values for the observables.  
Half the difference between the two extremes was used as the systematic 
uncertainty associated with the diagonal cut efficiency.

We calculated the systematic error due to the efficiency correction of
the $J/\psi, \pi^0$, and $\gamma$-conversion vetoes by using the ``hardest'' 
and ``softest'' primary-spectrum variations, as in the determination of 
the diagonal cut systematic.  We then took as the error half the
difference between the ``hardest'' and ``softest'' variations, plus 10\%
of itself.  This extra 10\% on the error was to account for the fact that
we only varied about 90\% of the primary spectrum when we reweighted the
unlike-sign spectrum.  Because of mixing, the other 10\% of the primary 
electrons appeared in the like-sign spectrum.  

\subsection{$\Delta(p)$ and $B^0\overline{B^0}$ Mixing}
\label{sec:syst_mixing_and_delta}

The factor $\Delta(p)$ accounts for the difference between the 
secondary-electron spectra in charged and neutral $B$ decays,
as described in Sec.~\ref{sec:spectrum_method}.  The systematic
uncertainty assigned to this was taken to be half of the difference 
between results obtained from Eqs.~(\ref{eq:primary}) and
(\ref{eq:secondary})  
with the $\Delta(p)$ determined in our Monte Carlo study (standard case)
and those obtained by taking with $\Delta(p) = 1$ (no correction).  

The uncertainty on the mixing parameter $\chi$ was determined from
relevant input data, as is described in Sec.~\ref{sec:spectrum_method}.
The effect on measured quantites was determined by solving for the
spectra with values of $\chi$ that were shifted up and down by
$1\sigma$.

\subsection{Cross-Checks}
\label{sec:cross_checks}
We also performed several cross-checks of our results to test all aspects of the 
analysis procedure and to verify that there were no biases in the determination
of the $B$ semileptonic branching fraction and electron-energy moments.  A $B {\bar B}$ 
Monte Carlo sample with known semileptonic branching fraction and spectral shape
was subjected to nearly the full analysis procedure.  Results obtained were 
consistent with inputs and generator-level quantities to within
statistical errors.

Other cross-checks involved subdividing the data sample in various ways to
demonstrate that there were no unexpected dependences in the results.  No 
statistically significant differences were found between the subsample with
electron tags and that with muon tags, between positively charged and negatively 
charged tags, between low-momentum ($<1.75~$GeV/$c$) and high-momentum 
($>1.75~$GeV/$c$) tags, or between the data samples collected before and
after the detector upgrade.  More details on these cross-checks can be
found in Ref.~\cite{stepaniak_thesis}. 

\section{$B$ Semileptonic Branching Fraction}
\label{sec:SLBR}
Integrating the measured primary spectrum in Fig.~\ref{fg:solved} between
0.6~GeV/$c$ and 2.6~GeV/$c$ gives the partial branching fraction
${\mathcal B}(B \rightarrow X e \nu, p>0.6~{\rm GeV}/c)=(10.21 \pm 0.08 \pm 0.22$)\%,
where the first uncertainty is statistical and the second is the systematic
uncertainty associated with measurement of the electron spectrum 
(Sec.~\ref{sec:spectrum_systematics}).  This result is almost completely
free of model dependence.  To extract the full semileptonic branching 
fraction, it is necessary to correct for the undetected portion of the 
electron spectrum below the low-momentum limit of 0.6~GeV/$c$.

To determine this fraction, we fitted the measured primary spectrum with a mixture of 
predicted spectra for the decay modes $B \rightarrow D e \nu$, 
$B \rightarrow D^* e \nu$, $B \rightarrow  D^{**} e \nu$, 
$B \rightarrow D X e \nu$, and charmless decays $B \rightarrow X_u e \nu$.  
All spectra were obtained from full GEANT~\cite{GEANT} simulations of
$B\bar{B}$ events and included 
electroweak radiative corrections as described by the PHOTOS 
algorithm \cite{Barberio:1994qi}.  The decays $B \rightarrow D^* e \nu$ were 
generated according to HQET with CLEO-measured form-factor parameters
\cite{Duboscq:1996mv}.
$B \rightarrow D e \nu$ decays were generated with the ISGW2 \cite{Scora:1995ty} 
model, and then reweighted to correspond to HQET with the form factor $\rho^2$ as 
measured by CLEO \cite{Athanas:1997eg}.  These $B \rightarrow D e \nu$ and 
$B \rightarrow D^* e \nu$ components of the fit were constrained 
to be within $\pm 2\sigma$ of the measured exclusive branching 
fractions \cite{PDG_2002}.  The third fit component, denoted 
$B \rightarrow D^{**} e \nu$, represented a mixture of decays to
higher-mass charmed mesons as described by ISGW2 \cite{Scora:1995ty}.
The fourth component was nonresonant $B \rightarrow D X e\nu$ 
as described by the model of Goity and Roberts \cite{Goity:1995xn}.  These last two 
were constrained in the fit only to the extent that they were not allowed to be 
negative.  The final component was the charmless decays $B \rightarrow
X_u \ell \nu$ modeled with 
a hybrid inclusive/exclusive generator developed by CLEO.  This model was built on 
the inclusive description of $B \rightarrow X_u \ell \nu$ developed by DeFazio and 
Neubert \cite{DeFazio:1999sv}, with shape-function parameters determined by 
fitting CLEO's inclusively measured $B \rightarrow X_s \gamma$ energy spectrum 
\cite{Chen:2001fj}.  For all final states with hadronic masses up to that of the 
$\rho(1450)$, exclusive final states, as described by the ISGW2 model \cite{Scora:1995ty}, 
were substituted.  The normalization
of the $B \rightarrow X_u e \nu$ component was fixed by the partial branching 
fraction in the 2.2-2.6~GeV/$c$ momentum region measured by CLEO \cite{Bornheim:2002du}.

The fit performed over $0.6<p_e<2.6$ GeV/$c$ according to these
specifications gave a $\chi^2$ of 34.5 for 
38 degrees of freedom, although it is noteworthy that the $B \rightarrow D e \nu$ and 
$B \rightarrow D^* e \nu$ branching fractions were pinned at their $+2 \sigma$ limits.
For this fit the fraction of the semileptonic decay spectrum below 600
MeV/$c$ was 0.064.

We assessed the systematic uncertainty in this estimate by performing a large number 
of variations of the standard fit.  In each case we refitted with only one ingredient 
changed.  The difference between the standard value for the spectral fraction and that 
for the modified fit was recorded as the systematic uncertainty associated with that 
ingredient, and the overall systematic uncertainty was obtained by combining in
quadrature.

The variations considered included $\pm 1 \sigma$ variations in the form-factor
parameters for $B \rightarrow D e \nu$ and $B \rightarrow D^* e \nu$, extreme variations 
in the rates of the less well known $D^{**}$ and nonresonant components, variations
in the normalization of the fixed $B \rightarrow X_u e \nu$ component, a 30\% variation
in the electroweak radiative corrections applied to the spectra (the approximate
difference between PHOTOS and the calculation of Atwood and
Marciano~\cite{Atwood:1989em}), 
and variations in the momentum scale with which $B$-decay distributions were boosted
into the lab frame.

A persistent feature of the fits in the above
list was that they demanded branching fractions for $B \rightarrow D e \nu$ and 
$B \rightarrow D^* e \nu$ that were not in good agreement with world-average 
values \cite{PDG_2002}.  To address this we also fitted the spectrum with the 
$B \rightarrow D e \nu$ and $B \rightarrow D^* e \nu$ branching fractions fixed 
to their PDG 2002 values, with the other $B \rightarrow X_c e \nu$ components left free.
The result was a very poor fit to the spectrum ($\chi^2$=85.5/38~d.o.f) and an
undetected spectral fraction of 0.070.  Even though this case was strongly disfavored 
by the measured electron spectrum, we included it in assessing the
systematic uncertainty.

Dividing the measured partial branching by the above-determined fraction of the
$B$ semileptonic momentum spectrum above 0.6~GeV/$c$ of $0.936 \pm 0.006$ gives
the total $B$ semileptonic branching ratio:

\begin{equation}
\mathcal{B}(B \rightarrow X e \nu) = (10.91 \pm 0.09 \pm 0.24)\%.
\end{equation}

The first uncertainty is statistical and the second is systematic.  The computation
of the systematic uncertainty is broken down in Table~\ref{tb:br_syst}.  
\begin{table}[hptb]
\begin{center}
\begin{tabular}{|l|c|}
\hline
Source                   & $\Delta\mathcal{B}_{SL}(\%)$ \\ \hline\hline
$J/\psi$            	 & 0.003 \\ \hline 
$\pi^0$             	 & 0.006 \\ \hline 
$\gamma$            	 & 0.023 \\ \hline 
Same $B$ secondaries	 & 0.052 \\ \hline 
Upper Vertex $D_s$  	 & 0.091 \\ \hline 
Upper Vertex $D$    	 & 0.065 \\ \hline 
$\tau$              	 & 0.041 \\ \hline 
$\psi(2S)$          	 & 0.005 \\ \hline 
Other Backgrounds      	 & 0.003 \\ \hline 
Tags from Secondaries	 & 0.014 \\ \hline
Electron Identification  & 0.113 \\ \hline 
Mixing Parameter    	 & 0.035 \\ \hline 
Continuum Subtraction	 & 0.028 \\ \hline 
Track Quality Efficiency    	 & 0.001 \\ \hline 
Diagonal Cut Efficiency   	 & 0.008 \\ \hline 
Veto Efficiency           	 & 0.006 \\ \hline 
Muon Fake Rate      	 & 0.001 \\ \hline 
$\Delta(p)$         	 & 0.021 \\ \hline 
Event Selection Ratio	 & 0.128 \\ \hline 
Fit Extrapolation   	 & 0.078 \\ \hline 
\hline
Total                    & 0.236 \\ \hline
\end{tabular}
\caption{\label{tb:br_syst}{Breakdown of systematic errors on
    $\mathcal{B}_{SL}$.}}
\end{center}
\end{table}

\section{Moments of the Electron-Energy Distribution}
\label{sec:moments}
Following the notation of Bauer {\it et al.} \cite{Bauer:2002sh}, we
define the electron-energy moments as follows:
\begin{equation}
R[n, E_{\ell_1}, m, E_{\ell_2}] = 
\frac{\int_{E_{\ell_1}}^{E_{\ell}^{max}}
E_{\ell}^n\frac{d\Gamma}{dE_{\ell}}dE_{\ell}}
{\int_{E_{\ell_2}}^{E_{\ell}^{max}}
E_{\ell}^m\frac{d\Gamma}{dE_{\ell}}dE_{\ell}},
\label{eq:moments}
\end{equation}
where $E_{\ell}^{max} = 2.5$~GeV.  For convenience, we denote $R[1,
E_{min}, 0, E_{min}]$ and $R[2, E_{min}, 0, E_{min}]$, as 
$\langle E_\ell\rangle $ and $\langle E_\ell^2\rangle $, with $E_{min}$
(in GeV) as a subscript when 
necessary.  We also use the spread of the spectrum, 
$\langle E_\ell^2 - \langle E_\ell\rangle ^2\rangle$ as an alternative to the second
moment, as it is less strongly correlated with $\langle E_\ell\rangle $
than $\langle E_\ell^2\rangle $. 
 
The moments computed theoretically are for the ``heavy-to-heavy'' decay
$B \rightarrow X_c \ell \nu$, while our spectrum and branching fraction
measurements included all semileptonic decays.  Before computing the energy 
moments we therefore subtracted the small contribution of 
$B \rightarrow X_u \ell \nu$ decays. The momentum spectrum for these
decays was generated with the hybrid inclusive/exclusive model 
described in Sec.~\ref{sec:SLBR} and the normalization was obtained
from the CLEO inclusive end-point measurement \cite{Bornheim:2002du}. 
To assess the systematic uncertainty associated with this subtraction,
we varied both the normalization and the shape of the 
$B \rightarrow X_u \ell \nu$ component.  CLEO's inclusive and exclusive 
\cite{Athar:2003yg} $B \rightarrow X_u \ell \nu$ measurements have shown 
that the proportion of the end-point ($2.2 - 2.6$~GeV/$c$) spectrum that
is due to $B \rightarrow \pi / \rho / \eta / \omega \ell \nu$
is approximately 55\%.  This has been used for the central value
in the hybrid model, and variations of $\pm 30\%$ in the exclusive
component were used to assess the sensitivity to the spectral
shape.  The normalization was varied up and down by one standard
deviation, using the combined statistical and systematic uncertainty of
the end-point measurement.

After subtracting the $B \rightarrow X_u \ell \nu$ from the spectrum
of Fig.~\ref{fg:solved}, we obtained the final $B \rightarrow X_c \ell \nu$
spectrum shown in Fig.~\ref{fg:final_spectrum}.  
\begin{figure}[!ht]
\centerline{
\epsfxsize=15 cm
\epsffile{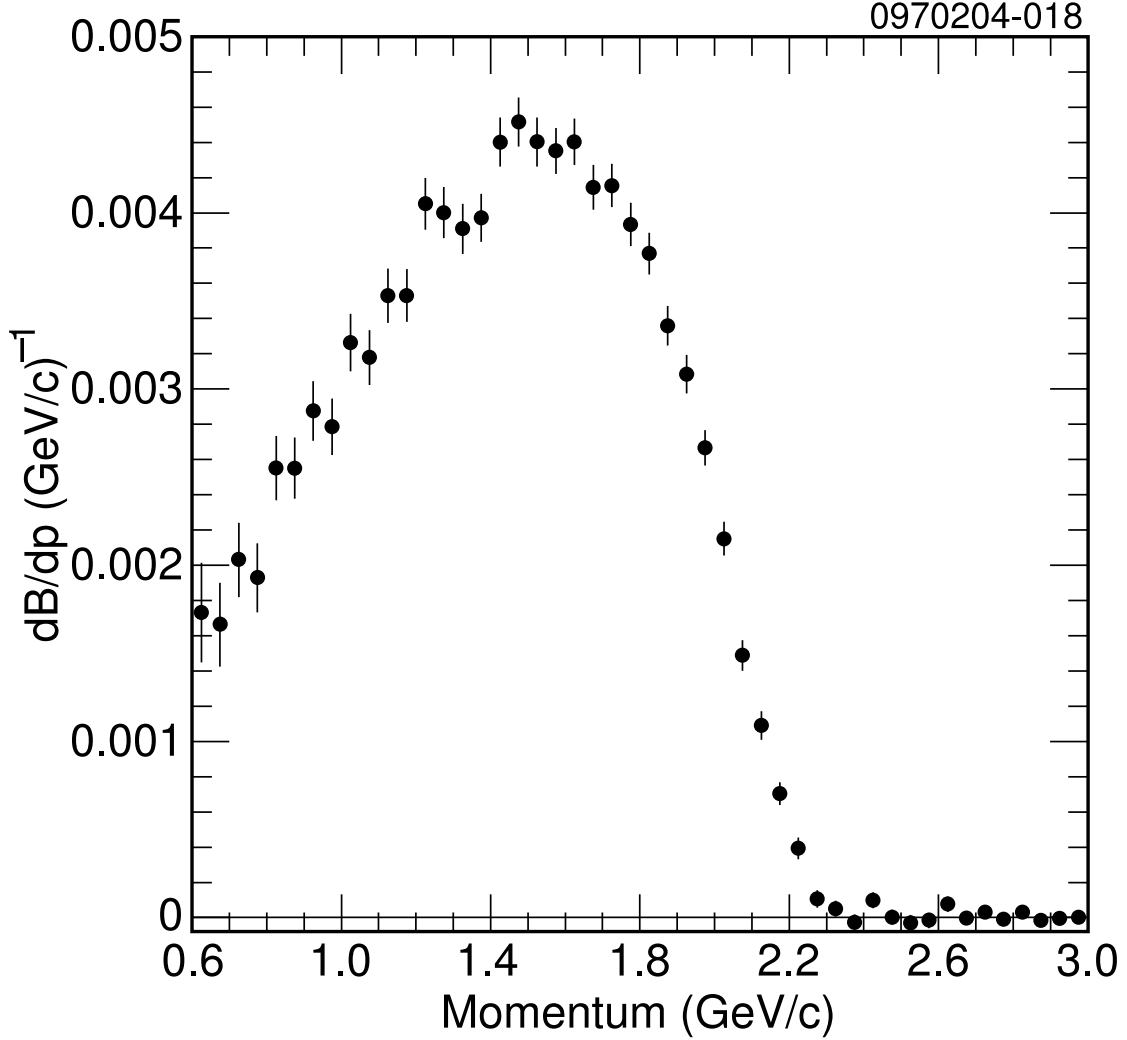}
}
\caption[fg:final_spectrum]
{\label{fg:final_spectrum} The final $B \rightarrow X_c \ell \nu$
spectrum.}
\end{figure}
From this spectrum we computed ``raw'' moments by direct integration.  These 
moments required two corrections before they could be interpreted with the 
theoretical expressions.  Because our moments were measured in the $\Upsilon(4S)$ 
rest frame, it was necessary to correct for the boost of the spectrum from the 
$B$ rest frame, where theoretical predictions are calculated.  This is a
very straightforward 
incorporation of the approximately 300~MeV/$c$ momentum of $B$ mesons produced
from an $\Upsilon(4S)$ decay at rest.  It could be done quite well analytically, 
although we performed it using Monte Carlo simulations that included the precise
beam-energy distribution of our data sample.  Using Monte Carlo samples, the value 
of each moment was computed in the $B$ and $\Upsilon(4S)$ rest frames and the difference 
was taken as an additive correction to be applied to the moment.  The sensitivity
to the momentum scale was explored by reweighting the spectra in $B$ momentum and
recomputing.  The sensitivity to decay mode and model was shown to be
negligible.   For $\langle E_{\ell}\rangle_{0.6}$ this correction is
$(-2.4 \pm 0.2)$ MeV.

The second correction was for electroweak final-state radiation, which is not 
generally included in the theoretical expressions.  Again, an additive correction
was obtained, in this case using the PHOTOS
algorithm~\cite{Barberio:1994qi} to generate spectra for different modes and models
and computing the differences in moment values with and without the correction.  
For comparison and assessment of the systematic uncertainty associated 
with this correction, we also used the calculation of Atwood and Marciano 
\cite{Atwood:1989em}.  The systematic uncertainty due to 
the electroweak correction was taken to be the difference between 
Atwood and Marciano and PHOTOS.  
For $\langle E_{\ell} \rangle_{0.6}$ this correction is $(+16.8 \pm
6.0)$ MeV.  This is the largest systematic error in the moments
measurement. 

From our final spectrum, and after the two corrections described above were
applied, we obtained values for electron-energy moments with minimum energies
between 0.6~GeV and 1.5~GeV.  These are given in
Table~\ref{tb:moments}.  Note that these numbers are highly correlated.
As a cross-check of our procedure for
extracting the moments, we also computed them from the $B \to X_c \ell
\nu$ spectra obtained with the fits to Monte Carlo-predicted spectra as
described in Sec.~\ref{sec:SLBR}.  Consistent results were obtained in
all cases. 
\begin{table}
\begin{center}
\begin{tabular}{|c|c|c|c|}
\hline
$E_{min}$ & {\bf $\langle E_\ell\rangle $ (GeV)} & {\bf $\langle E_\ell^2\rangle $ (GeV$^2$)} & {\bf $\langle E_\ell^2-\langle E_\ell\rangle ^2\rangle $ (GeV$^2$)} \\ \hline\hline
{\bf 0.6} & $1.4261 \pm 0.0043 \pm 0.0105 $ & $2.1856 \pm 0.0112 \pm 0.0271 $ & $0.1526 \pm 0.0021 \pm 0.0031 $ \\\hline 
{\bf 0.7} & $1.4509 \pm 0.0035 \pm 0.0079 $ & $2.2419 \pm 0.0097 \pm 0.0216 $ & $0.1374 \pm 0.0015 \pm 0.0018 $ \\\hline
{\bf 0.8} & $1.4779 \pm 0.0031 \pm 0.0061 $ & $2.3066 \pm 0.0090 \pm 0.0177 $ & $0.1228 \pm 0.0013 \pm 0.0012 $ \\\hline
{\bf 0.9} & $1.5119 \pm 0.0028 \pm 0.0047 $ & $2.3923 \pm 0.0085 \pm 0.0144 $ & $0.1068 \pm 0.0011 \pm 0.0010 $ \\\hline
{\bf 1.0} & $1.5483 \pm 0.0026 \pm 0.0039 $ & $2.4890 \pm 0.0082 \pm 0.0127 $ & $0.0918 \pm 0.0010 \pm 0.0011 $ \\\hline
{\bf 1.1} & $1.5884 \pm 0.0024 \pm 0.0033 $ & $2.6003 \pm 0.0080 \pm 0.0111 $ & $0.0775 \pm 0.0009 \pm 0.0012 $ \\\hline
{\bf 1.2} & $1.6315 \pm 0.0023 \pm 0.0031 $ & $2.7259 \pm 0.0078 \pm 0.0109 $ & $0.0642 \pm 0.0009 \pm 0.0012 $ \\\hline
{\bf 1.3} & $1.6794 \pm 0.0022 \pm 0.0029 $ & $2.8720 \pm 0.0078 \pm 0.0106 $ & $0.0516 \pm 0.0008 \pm 0.0011 $ \\\hline
{\bf 1.4} & $1.7256 \pm 0.0021 \pm 0.0030 $ & $3.0192 \pm 0.0079 \pm 0.0112 $ & $0.0413 \pm 0.0008 \pm 0.0010 $ \\\hline
{\bf 1.5} & $1.7792 \pm 0.0021 \pm 0.0027 $ & $3.1972 \pm 0.0081 \pm 0.0107 $ & $0.0316 \pm 0.0008 \pm 0.0010 $ \\\hline
\end{tabular}
\caption{\label{tb:moments} {Electron-energy moments for various minimum lepton-energy cuts $E_{min}$.}}
\end{center}
\end{table}

Systematic uncertainties in the moment values were assessed with the techniques
described in Sec.~\ref{sec:spectrum_systematics} (background and efficiency corrections) 
and earlier in this section (moment extraction).  To provide a concrete illustration,
the mean energy for the full measured spectrum is
$\langle E_{\ell}\rangle_{0.6} = (1.4261 \pm 0.0043 \pm 0.0105)$~GeV, where the
first error is statistical and the second is systematic.  The largest
sources of systematic uncertainty for this moment are the electroweak
radiative correction ($\pm 0.0060$), upper-vertex charm background
correction ($\pm 0.0059$), and electron identification ($\pm 0.0046$). 
All of these, and the total systematic uncertainty, diminish with
increasing minimum-energy cut, as shown in Table~\ref{tb:moments}. 

\section{Interpretation and Conclusions}
\label{sec:conclusions}
In this paper we have presented a new measurement of the inclusive momentum
spectrum for semileptonic $B$-meson decays using events with a high-momentum
lepton tag and a signal electron in the full data sample collected with the 
CLEO~II detector.  Improvements in the understanding of background processes 
and optimized electron-identification procedures have resulted in significant 
improvements in systematic uncertainties relative to the previous CLEO measurement
\cite{Barish:1995cx}, which this analysis supersedes.  We have used
the normalization of the measured 
spectrum and an extrapolation for $0 < E_\ell < 0.6$ GeV based on a detailed
model calculation constrained by data to obtain a new measurement of the
$B$ semileptonic branching  
fraction, $\mathcal{B}(B \rightarrow X e \nu) = (10.91 \pm 0.09 \pm 0.24)\%$. This
result is in excellent agreement with other recent measurements at the $\Upsilon(4S)$ 
\cite{Aubert:2002uf, Abe:2002du} and has better overall precision.  These
results have diminished the 
level of disagreement between measurements made at the $\Upsilon(4S)$ and those
from $Z^0$ decays~\cite{Albrecht:1993pu}.  While still somewhat lower
than theoretical predictions, the measured $B$ semileptonic branching
fraction is now less in conflict~\cite{Bigi:1994fm} with them than was previously
the case.

We have also used our measured spectrum to determine the moments of electron energy
in semileptonic $B$ decays with minimum energies ranging from 0.6~GeV to 1.5~GeV
(Table~\ref{tb:moments}).
Our measured value for the mean energy with $E_{min}=1.5$~GeV/$c$, 
$\langle E_{\ell}\rangle_{1.5} = (1.7792 \pm 0.0021 \pm 0.0026)$~GeV, is 
in good agreement with the previous CLEO measurement of this
quantity~\cite{Mahmood:2002tt}, 
$(1.7810 \pm 0.0007 \pm 0.0009)$ GeV.  The earlier measurement was more precise because 
it used the entire inclusive spectrum for semileptonic $B$ decays, 
without a lepton-tag requirement.  That technique does not allow for measurements
with smaller values of $E_{min}$, however, because of the large contribution of 
secondary charm decays.  While electron-energy moments were not presented for the 
previous CLEO lepton-tagged measurement of $\mathcal{B}(B \to X e \nu)$ 
\cite{Barish:1995cx}, we note that moment values computed from
fits to that spectrum are consistent with the current measurements.

Measurements of moments of different quantities and with sensitivity to 
different regions of phase space provide an ideal opportunity to test the
description of inclusive $B$ decays provided by the HQET/OPE methodology.
Using this approach, theorists have derived
expressions~\cite{Bauer:2002sh} for many inclusive 
properties of $B$ decays, including the moments of the 
lepton energy and recoil hadronic mass in $B \rightarrow X_c \ell \nu$ 
and of the photon energy in  $B \rightarrow X_s \gamma$.  The physical 
observables are expressed as expansions in $\Lambda_{QCD}/M_B$ and new 
parameters emerge at each order: ${\bar \Lambda}$ at order $\Lambda_{QCD}/M_B$, 
$\lambda_1$ and $\lambda_2$ at order $\Lambda_{QCD}^2/M_B^2$, and six parameters 
($\rho_1$, $\rho_2$, ${\mathcal T}_1$, ${\mathcal T}_2$, ${\mathcal T}_3$, 
${\mathcal T}_4$) at order $\Lambda_{QCD}^3/M_B^3$~\cite{Gremm:1997df}.

Previous CLEO moments measurements~\cite{Chen:2001fj,
Cronin-Hennessy:2001fk, Mahmood:2002tt} have been interpreted with
theoretical expansions in the pole-mass 
scheme to order $\beta_0(\alpha_s/\pi)^2$ in the perturbative and
$\Lambda_{QCD}^3/M_B^3$ in the nonperturbative expansion.  The six
third-order parameters were fixed in fitting the data, and fluctuated
within bounds determined by dimensional arguments~\cite{Gremm:1997df}
for assessment of the uncertainty.  A combined fit to the data gave
${\bar \Lambda}=(0.39 \pm 0.14)$~GeV and $\lambda_1=(-0.25 \pm
0.15)$~GeV$^2$, where the uncertainties are dominated by
theory~\cite{Mahmood:2002tt}. 

\begin{figure}[htp]
\centerline{
\begin{minipage}{8.5 cm}
\epsfxsize=8.5 cm
\epsfysize=8.5 cm
\epsffile{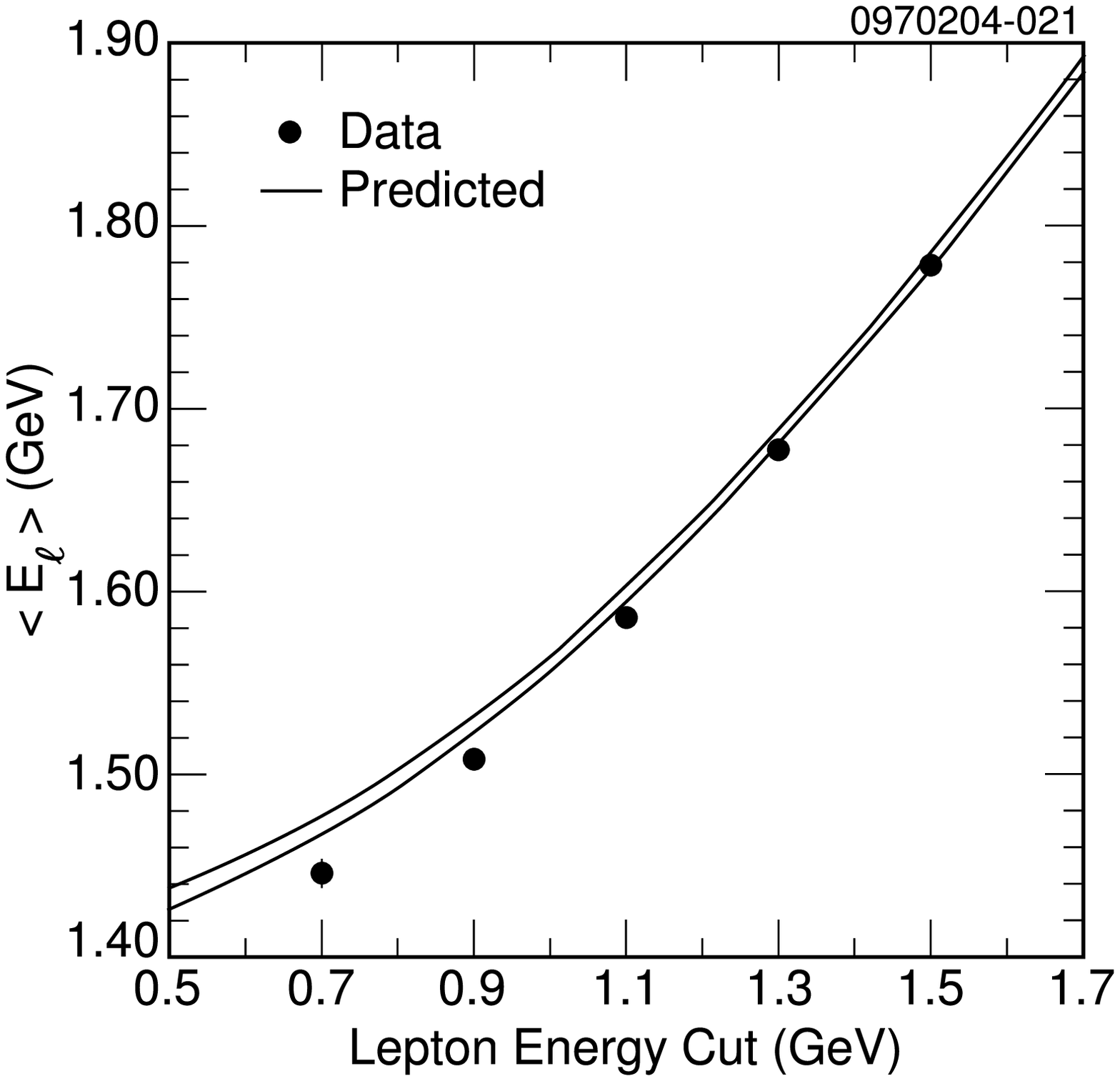}
\end{minipage}
\begin{minipage}{8.5 cm}
\epsfxsize=8.5 cm
\epsfysize=8.5 cm
\epsffile{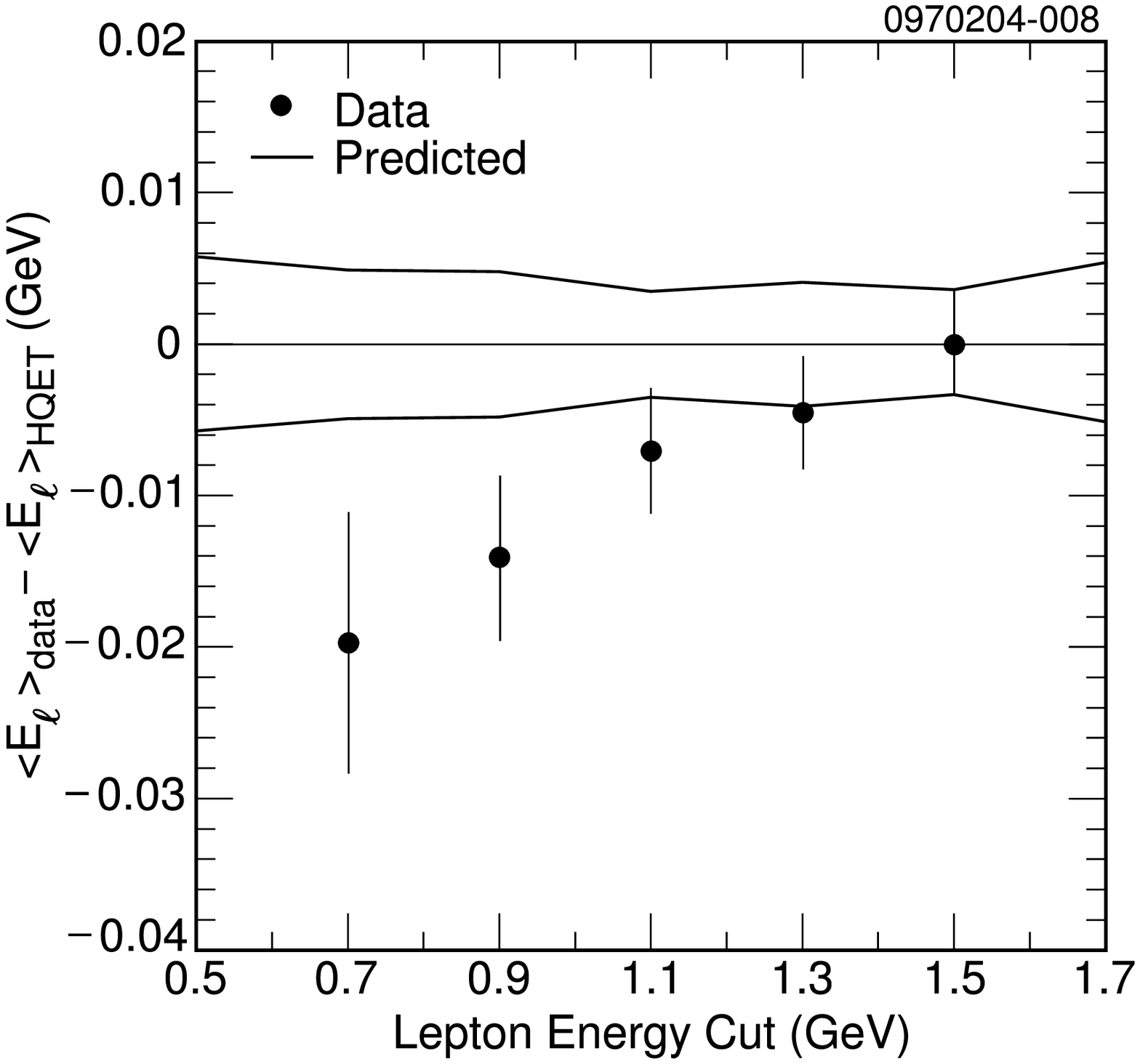}
\end{minipage}
}
\caption[fg:first_moment_vs_HQET]
{\label{fg:first_moment_vs_HQET} {\bf Left:} $\langle
  E_{\ell}\rangle$ as a function of $E_{min}$.  The points
  are data and the band is the $\pm 1 \sigma$ prediction in the
  pole-mass scheme~\cite{Bauer:2002sh}.  {\bf Right:} $\langle 
  E_\ell\rangle_{data} - \langle E_\ell\rangle_{HQET}$ as a
  function of $E_{min}$.  The points are the data from
  Table~\ref{tb:moments} and the band
  is the $\pm 1 \sigma$ prediction in the pole-mass scheme.  Inputs for
  these plots were set by the first photon energy moment of $b \to
  s\gamma$~\cite{Chen:2001fj} and $\langle E_{\ell}\rangle_{1.5}$.} 
\end{figure}

The plots in Fig.~\ref{fg:first_moment_vs_HQET} show our measured values of
$\langle E_\ell\rangle$ as a function of the minimum lepton energy cut
and the HQET/OPE predictions for the electron-energy moments in the pole
mass scheme provided by \cite{Bauer:2002sh}.  The plot on the left shows the
measurements and the prediction, while the plot on the right shows the
difference between the measurements and the prediction.  The values for
$\bar{\Lambda}$ and $\lambda_1$ are constrained by the first
photon-energy moment of the $b\to s\gamma$ spectrum~\cite{Chen:2001fj} 
and  our measurement of $\langle
E_{\ell}\rangle_{1.5}$.  The third-order parameters ${\mathcal T}_{1-4}$ were
taken to be to $(0 \pm 0.5$ GeV$)^3$.  The parameter $\rho_1$ was taken to
be ($0.0625 \pm 0.0625$)~GeV$^3$~\cite{Gremm:1997df}, and $\rho_2$ is constrained
by $B^*-B$ and $D^*-D$ mass splittings~\cite{Bauer:2002sh}.  The error bars 
on the data points represent the combined statistical and systematic
uncertainties of the measurements.  There is substantial correlation
among the data values for the different $E_{min}$ cases.  The width of
the band is set by the uncertainty in the measurements of ${\bar
  \Lambda}$ and $\lambda_1$, variation of the third-order
expansion parameters, and variation of the perturbative QCD
corrections. 

As can be seen in Fig.~\ref{fg:first_moment_vs_HQET}, there is an
increasing disagreement as $E_{min}$ is reduced between the measured
mean energy and the value extrapolated with HQET.  We note again 
that these results have been obtained by using the PHOTOS algorithm 
\cite{Barberio:1994qi} to correct for final-state radiation.  
There is considerable uncertainty in this correction, and if the
prescription of Atwood and Marciano \cite{Atwood:1989em} 
were instead used, the disagreement between our measurement and the 
HQET computation would be increased by 25\%.  The difference between
these two computations is the largest contribution to the systematic
uncertainty in the measurement of the mean energy.

\begin{figure}[htp]
\centerline{
\epsfxsize=14 cm
\epsffile{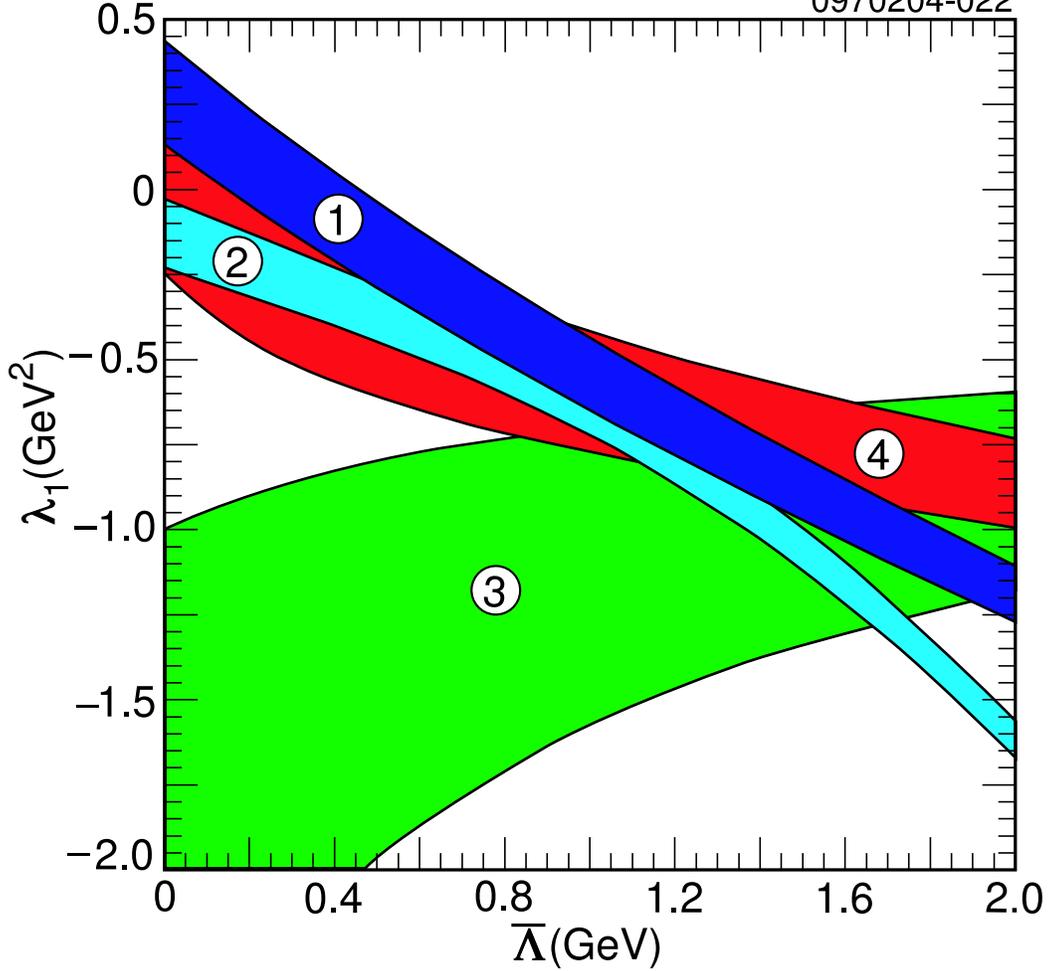}
}
\caption[fg:bands]
{\label{fg:bands} Bands in the ${\bar \Lambda}-\lambda_1$
  plane from $\langle E_{\ell}\rangle$ with $E_{\ell} > 0.7$ GeV
  (band 1),
  $\langle E_\ell^2-\langle E_\ell\rangle ^2\rangle$ with $E_{\ell} >
  0.7$ GeV (band 2), $\langle E_{\ell}\rangle_{1.5} - \langle
  E_{\ell}\rangle_{0.7}$ (band 3), and $(\langle E_\ell^2-\langle
  E_\ell\rangle^2\rangle)_{0.7} - (\langle E_\ell^2-\langle
  E_\ell\rangle^2\rangle)_{1.5}$ (band 4).  The widths of the bands reflect the
  combined experimental and theoretical $1\sigma$ uncertainties.  These bands were
  calculated in the pole mass scheme~\cite{Bauer:2002sh}.}
\end{figure}

Fig.~\ref{fg:bands} shows four bands in the ${\bar \Lambda}-\lambda_1$ space.
Along with the standard bands for $\langle E_{\ell}\rangle_{0.7}$ and
$(\langle E_\ell^2-\langle E_\ell\rangle^2\rangle)_{0.7}$, we show
bands for the difference of the mean $\langle E_{\ell}\rangle_{1.5} -
\langle E_{\ell}\rangle_{0.7}$ and the difference in the variance
$(\langle E_\ell^2-\langle E_\ell\rangle^2\rangle)_{0.7} - (\langle
E_\ell^2-\langle E_\ell\rangle^2\rangle)_{1.5}$ to isolate the
information that is independent of the measurements of the moments with
$E_{\ell} > 1.5$ GeV.  The width of the bands indicates the combined
experimental and theoretical uncertainties.  As can be seen, the
variance (band 2) and the difference in the variances (band 4) are
compatible with other measurements~\cite{Mahmood:2002tt}, whereas the
difference in the means (band 3) is the predominant source of disagreement
between data and theory. 

There are several possible explanations for the observed inconsistency 
within HQET of the parameters extracted from our different energy-moment 
measurements.  In light of the sizable disagreement between the PHOTOS 
and Atwood/Marciano treatments of electroweak radiation, we cannot exclude 
an error in this correction that is outside of the quoted systematic 
uncertainty, although it seems unlikely.  Possible theoretical
explanations include problems with the specific HQET/OPE implementations 
that we have used, incorrect assumptions about the unknown third-order 
parameters, and problems with the underlying assumptions, such as 
quark-hadron duality.  A comprehensive fit, including correlations, of 
all published CLEO moments \cite{Chen:2001fj, Cronin-Hennessy:2001fk, 
Mahmood:2002tt}, the electron-energy moments in this paper, and new 
measurements of the recoil hadronic mass moments in 
$B \rightarrow X_c \ell \nu$ \cite{Lipeles_moments} is currently in 
preparation. By leaving parameters free at third order, this will 
determine if any of the HQET/OPE formulations, including the different 
mass schemes presented by Bauer {\it et al.} \cite{Bauer:2002sh} and the
kinetic mass scheme of Uraltsev {\it et al.}~\cite{Gambino:2004qm}, 
can accommodate all of the data.

During the final preparation of this paper, we learned of a preprint
from the BaBar collaboration reporting new measurements of the moments
of the electron-energy spectrum in semileptonic $B$ decays \cite{Aubert:2004td}.  The
BaBar results are based on an $\Upsilon(4S)$ sample with about five
times the integrated luminosity of our CLEO II data and are consistent
within quoted uncertainties with the measurements reported in this
paper.  The combined statistical and systematic uncertainties of the
BaBar results range from essentially identical to those of our
measurements (partial semileptonic branching fraction) to approximately
two thirds as large (first energy moments). 

\section{Acknowledgments}
\label{sec:Ack}

We gratefully acknowledge the efforts of the CESR staff in providing our
excellent data sample.  We thank C. Bauer, N. Uraltsev, A. Vainshtein, 
M. Voloshin, and P. Gambino for their useful discussions and
correspondence.  This work was supported by the National Science
Foundation, the U.S. Department of Energy, the Research Corporation, and
the Texas Advanced Research Program. 

\bibliography{sltag_revtex4}

\end{document}